\documentclass[journal]{IEEEtran}
\IEEEoverridecommandlockouts

\usepackage{multirow}     %
\usepackage{booktabs}     %
\usepackage{amsmath,amssymb,amsfonts}
\usepackage{rotating}
\usepackage{graphicx}
\usepackage{adjustbox}
\usepackage{longtable}
\usepackage{subcaption}
\usepackage{array}
\usepackage[table,xcdraw]{xcolor}
\newcolumntype{H}{>{\setbox0=\hbox\bgroup}c<{\egroup}}
\newcolumntype{Z}{>{\setbox0=\hbox\bgroup}c<{\egroup}@{\hspace*{-\tabcolsep}}}

\def\BibTeX{{\rm B\kern-.05em{\sc i\kern-.025em b}\kern-.08em
    T\kern-.1667em\lower.7ex\hbox{E}\kern-.125emX}}
\begin{document}

\bstctlcite{IEEEexample:BSTcontrol}
\title{IQNet: Image Quality Assessment Guided Just Noticeable Difference Prefiltering For Versatile Video Coding \\
}
\author{\IEEEauthorblockN{Yu-Han Sun, Chiang Lo-Hsuan Lee and Tian-Sheuan Chang, \textit{Senior Member, IEEE}}
\thanks{This work was supported by Novatek and the National Science and Technology Council, Taiwan, under Grant 111-2622-8-A49-018-SB, 110-2221-E-A49-148-MY3, and 110-2218-E-A49-015-MBK. The authors are with the Institute of Electronics, National Yang Ming Chiao Tung University, Hsinchu 30010, Taiwan (e-mail: mee10323.ep05@nctu.edu.tw, chiang0106.ee10@nycu.edu.tw, tschang@nycu.edu.tw) }%
\thanks{To be published in IEEE Open Journal of Circuits and Systems, Digital Object Identifier or DOI: 10.1109/OJCAS.2023.3344094}
}
\maketitle

\begin{abstract}%
Image prefiltering with just noticeable distortion (JND) improves coding efficiency in a visual lossless way by filtering the perceptually redundant information prior to compression. However, real JND cannot be well modeled with inaccurate masking equations in traditional approaches or image-level subject tests in deep learning approaches. Thus, this paper proposes a fine-grained JND prefiltering dataset guided by image quality assessment for accurate block-level JND modeling. The dataset is constructed from decoded images to include coding effects and is also perceptually enhanced with block overlap and edge preservation. Furthermore, based on this dataset, we propose a lightweight JND prefiltering network, IQNet, which can be applied directly to different quantization cases with the same model and only needs 3K parameters. The experimental results show that the proposed approach to Versatile Video Coding could yield maximum/average bitrate savings of 41\%/15\% and 53\%/19\% for all-intra and low-delay P configurations, respectively, with negligible subjective quality loss. Our method demonstrates higher perceptual quality and a model size that is an order of magnitude smaller than previous deep learning methods.

\end{abstract}


\section{Introduction}
\label{chapter:introduction}

Perceptual video coding (PVC) has attracted increasing attention as a means of improving coding efficiency by removing visual redundancy without perceptual loss beyond existing coding standards. PVC schemes generally develop just noticeable distortion (JND) models to determine the distortion that a human visual system can just notice.

JND models can be roughly classified into two types~\cite{survey}: pixel or frequency domain. Traditional methods ~\cite{7005446,7428849} build the pixel domain JND according to the luminance, contrast, and temporal masking of the human visual system, and add a contrast-sensitive function to adjust the above masking in a frequency-adaptive way for the frequency domain JND for better JND modeling. These JND models can be applied to input prefiltering before any encoder or quantization in the codec. This paper focuses on the input prefiltering that filters the original image based on the JND model and produces a visually lossless result compared to the original, which can be applied to any kind of codec. However, traditional methods generally do not consider coding effects in their models and could produce artifacts or have a limited improvement in coding efficiency, as indicated in~\cite{ki2018erjnd}.

Beyond traditional methods, with the popularity of deep learning, learning-based JND models have also gained significant attention in recent years for better modeling. For JND in quantization, two open datasets, VideoSet~\cite{videoset} and MCL-JCV~\cite{mcl}, are built based on subjective tests, which model video quality as a function of the quantization parameter (QP) to help train the JND model~\cite{bljnd,vwjnd}. \cite{vwjnd} combines spatial and temporal information based on convolutional neural networks (CNNs), which is only at the image level and therefore is not suitable for fine-grained processing as in this paper. \cite{9745537} conducts crowdsourced subject tests to build a large-scale JND image dataset, KonJND-1K. However, this is only for JPEG and BPG. 
\cite{MTJND} proposes a multitask deep learning framework to learn JND values, although without including the coding effects or evaluating its efficacy against the coding standards. \cite{mao2023transfer} propose a pixel-wise JND model that uses transfer learning, but only for JPEG. \cite{wang2023learning, wang2023surprise} proposed a linear regression based JND model for coding distortion to decide quantization parameters according to its entropy-based JND threshold.

In contrast, for JND on input prefiltering, there is no open dataset available for training, which becomes a challenge to develop a learning-based JND prefiltering model. To overcome this difficulty, \cite{unsuper} used unsupervised learning with SSIM and the total variation of the images to build their loss function for JND modeling. However, this JND model does not consider the coding effect. To address the problem of the dataset and include the coding effect, \cite{canf} proposed an end-to-end learning-based video compression network, and \cite{remodel} applied the surrogate codec with differentiable rate estimation to model the target codec for prefiltering network optimization. However, an end-to-end network with a surrogate codec could be difficult and time-consuming to train. Furthermore, if the surrogate codec is incorrectly modeled, the performance of the prefiltering networks could lead to poor results.
On the other hand, \cite{ki2018erjnd} conducted subjective tests that require time to build training data for their Energy-Reduced JND model (ERJND), which decreases the same magnitudes for all frequency coefficients in the 8x8 DCT domain. With this, they trained a CNN-based JND prefiltering model with a one-QP one-model approach for different QPs, called CNN-JNQD. They also extended ERJND and applied it to HDR video (HDR-JNDNet)~\cite{HDRNET}. However, their 8x8 block-based method could lead to blocking effects. The same magnitude reduction for all frequency bands does not fit the human visual system well. Their model has 14 layers ~\cite{HDRNET}, which will require high computational complexity for current high-definition video inputs. The one-QP one-model approach is not convenient for practical use. Last but not least, their scheme needs a subjective test to build the dataset, which is time-consuming and not scalable to larger numbers of data.

Addressing the aforementioned issues, this work presents an Image Quality Assessment (IQA)-guided Just Noticeable Difference (JND) dataset, along with a lightweight, learning-based JND prefiltering network. The contributions of this work are detailed below:

\begin{itemize}
\item \textbf{IQA-guided Fine-grained JND Prefiltering Dataset:} To address the absence of a fine-grained JND prefiltering dataset and to alleviate the high labor costs associated with human-conducted subjective tests, this paper constructs a fine-grained dataset for JND prefiltering under the guidance of no-reference IQA. The use of no-reference IQA allows for the derivation of JND values at a granular 64$\times$64 block level, rather than at the image level. Each block can undergo precise JND adjustments based on the quantitative IQA values. The IQA-centric approach enables the scalable creation of a larger dataset without incurring the labor costs of subjective testing. This dataset is further enriched by incorporating coding effects for JND modeling, which evaluates the quality of the reconstructed image after being encoded by Versatile Video Coding (VVC)\cite{VVC}. The image quality of the ground truth is further enhanced by deliberately retaining critical image details, leading to superior images and avoiding artifacts.

\item \textbf{Lightweight Attention-based JND Prefiltering:} The proposed JND prefiltering network, called IQNet, leverages a pixel attention mechanism, requiring only 3K model parameters to learn JND values without the need for explicit mask modeling. This streamlined model facilitates the application in real-time high-definition videos. Additionally, in contrast to the previous one-QP-one-model strategy, this network can accommodate various QPs with a single model, simplifying its deployment in practical scenarios.
\end{itemize}

The remainder of the paper is organized as follows: Section II details the construction of the training data. Section III proposes the JND prefiltering network, IQNet. Sections IV and V present the experimental results of the training data and IQNet, respectively. Finally, we conclude this paper in Section VI.

\section{IQA-Guided JND Dataset}
\subsection{Overview}

\begin{figure*}[tb]
\centering
\includegraphics[height=!,width=1.0\linewidth,keepaspectratio=true]{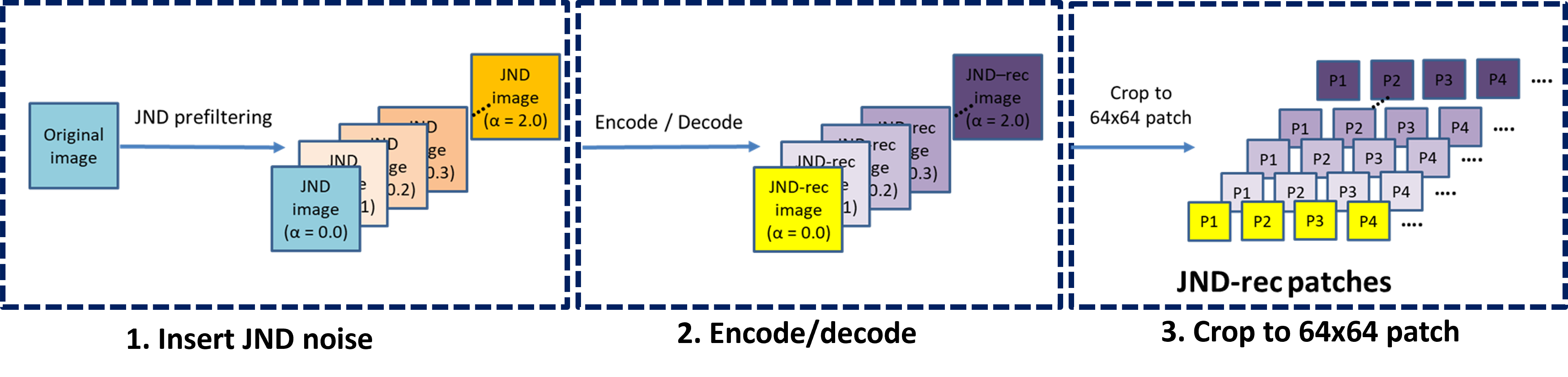}
\includegraphics[height=!,width=1.0\linewidth,keepaspectratio=true]{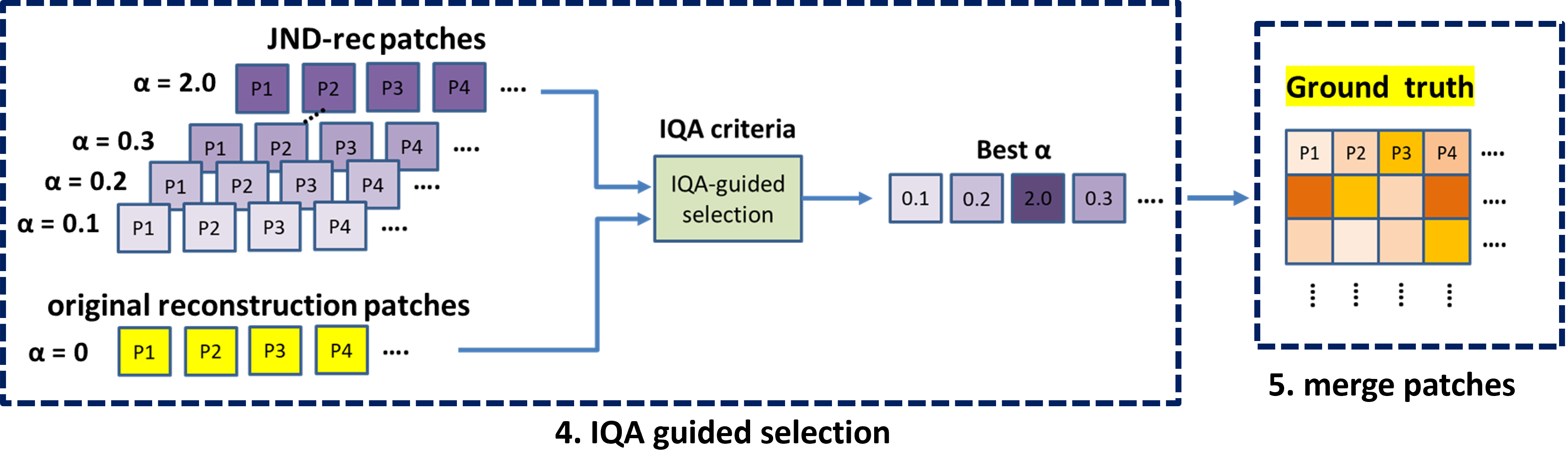}
\caption{The flow of training data generation.}
\label{groundtruth}
\end{figure*}

Fig.~\ref{groundtruth} illustrates the five steps of training data generation. For an image, we (1) apply JND prefiltering with different scales $\alpha$ to obtain possible JND candidates, (2) encode the prefiltered images and reconstruct them to include coding effects, (3) crop the reconstructed images to 64x64 patches for IQA evaluation, (4) select the best scale for each crop according to IQA, and (5) apply the selected JND to each patch of the original image and merge these patches to form the training image. The details of step (1) JND prefiltering and (4) IQA-guided selection will be described below.

\subsection{JND prefiltering for training data}

\begin{figure}[tb]
\centering
\includegraphics[height=!,width=1.0\linewidth,keepaspectratio=true]{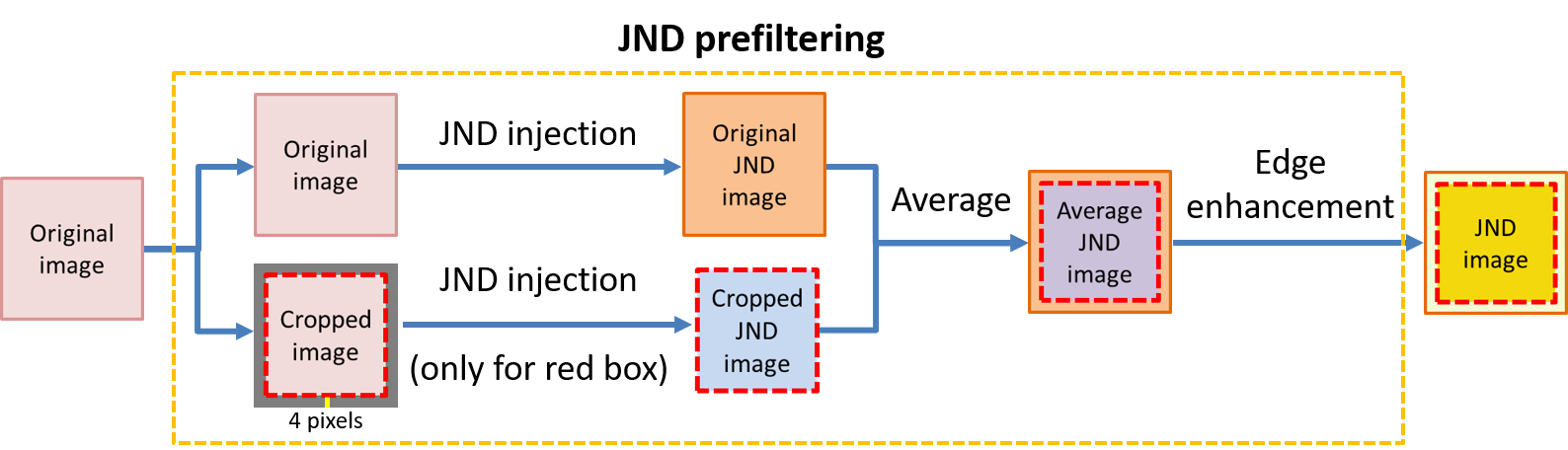}
\caption{JND prefiltering flow for training data.}
\label{prefiltering}
\end{figure}

Fig.~\ref{prefiltering} illustrates the flow of JND prefiltering for training data. In this process, we first inject different scaled JND values with DCT filtering into the original images and apply perceptual enhancement techniques such as boundary overlap and edge preservation to preserve important details, thereby producing JND-filtered images with high perceptual quality.

\subsubsection{JND Injection}

\begin{figure}[tb]
\centering
\includegraphics[height=!,width=1.0\linewidth,keepaspectratio=true]{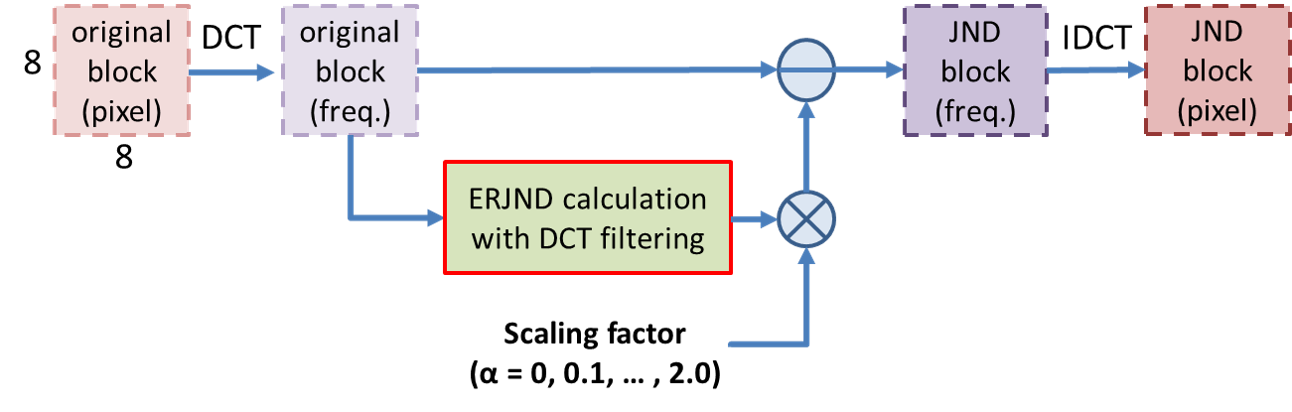}
\caption{JND injection}
\label{JND injection}
\end{figure}

To construct JND-filtered images, we first inject JND into the luminance of the original images. Our JND model is based on ERJND~\cite{ki2018erjnd}, but with additional perceptual enhancement. Following the ERJND process~\cite{ki2018erjnd}, we first partition the luminance of the original image into 8x8 blocks and apply DCT on each 8x8 block, since ERJND is applied to an 8x8 DCT domain, as shown in Fig.~\ref{JND injection}. Next, we inject ERJND noise with a different linear scaling factor $\alpha$ and DCT filtering (\textit{JND = ERJND$\times\alpha\times$DCT weighting}) to each 8x8 block to suppress the magnitudes of the DCT coefficients. After that, we convert the DCT coefficients to the pixel domain to obtain the luminance and combine it with the chrominance to obtain the JND block. The combination of all 8x8 JND blocks will be the resulting image.

\subsubsection{Frequency adaptive DCT filtering}

\begin{figure}[tb]
\centering
\includegraphics[height=!,width=1.0\linewidth,keepaspectratio=true]{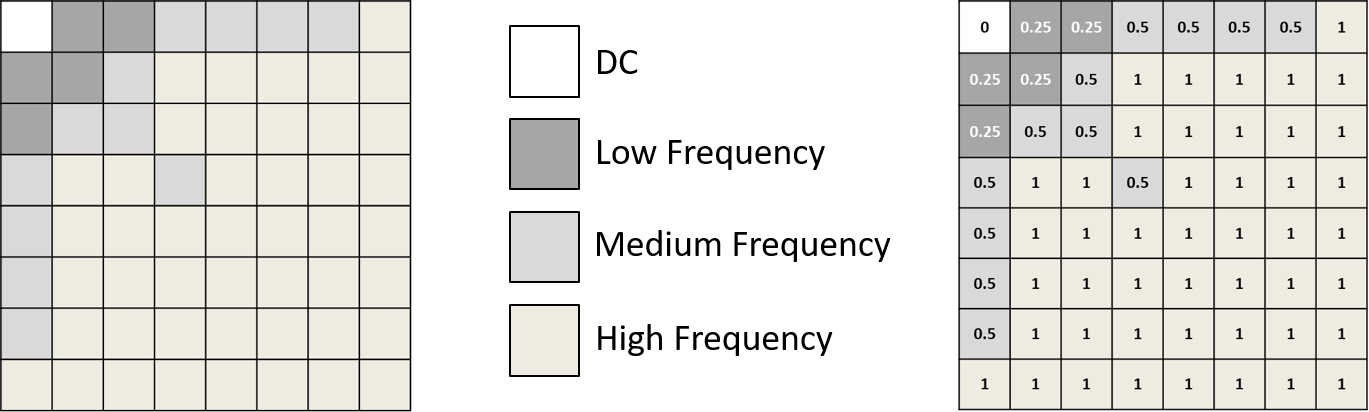}
\caption{DCT filter for an 8x8 DCT block.}
\label{DCT filter}
\end{figure}

The original ERJND does not consider the perceptual sensitivity to frequency characteristics, which could cause JND overfiltering in low-frequency areas and produce blurry images with a blocking effect. To solve this problem, we apply a DCT filter based on \cite{adaptive}\cite{anovel} to control the JND level in different frequency bands by multiplying the original ERJND value by filter weights as shown in Fig.~\ref{DCT filter}. In this figure, the weights will be smaller for the DC and low-frequency areas, since humans are more sensitive to these areas. In contrast, the weights will be larger for medium- and high-frequency areas due to lower perceptual sensitivity.

\subsubsection{Perceptual enhancement}

Beyond the above DCT filtering, we propose two perceptual enhancement methods to avoid the blocking effect and preserve more important details. A blocking effect will occur due to 8x8 blockwise JND injection when the scale factor $\alpha$ is large. To avoid this during IQA evaluation, we propose boundary overlap as shown in Fig.~\ref{prefiltering}. As shown in the figure, this overlapping is achieved by two JND injection paths for different parts of the same image. One is for the entire original image, and the other is a cropped JND image. This cropped JND image is generated by cropping four pixels on each side of the original image and applying the 8x8 JND filtering. Thus, the cropped JND images will have 8x8 blocks with a 4-pixel shift vertically and horizontally relative to the original image. When these two are combined, the cropped JND image will be averaged with the original JND image to reduce the blocking effect.

\begin{figure}[tb]
\centering
\includegraphics[height=!,width=1.0\linewidth,keepaspectratio=true]{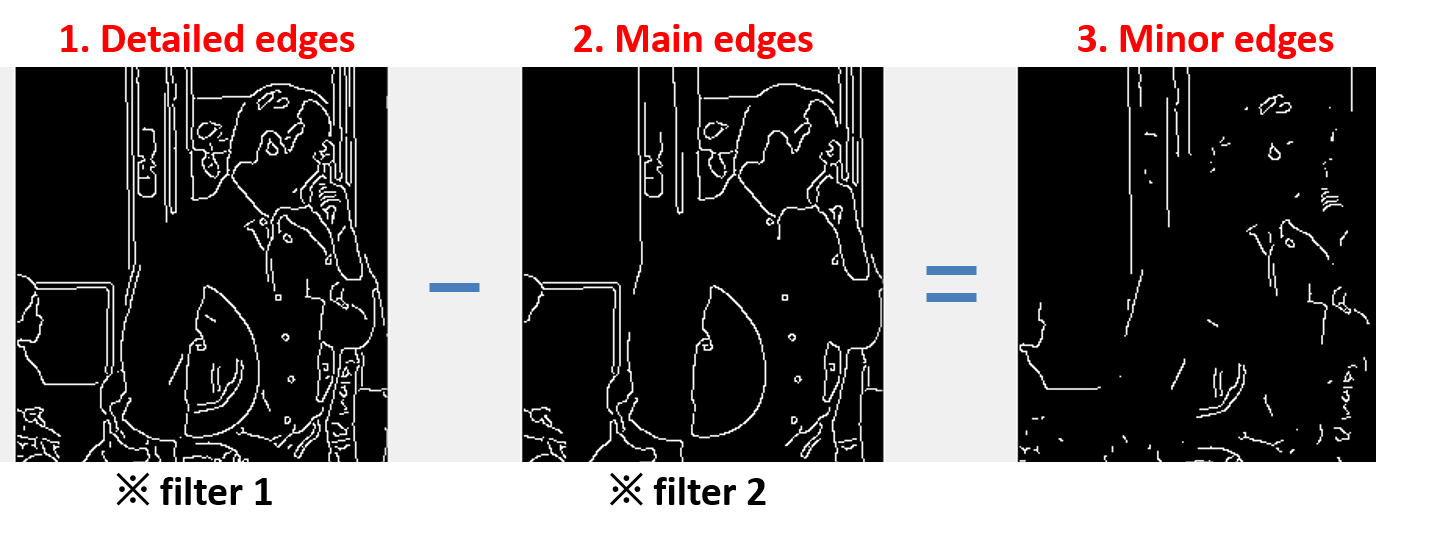}
\caption{Calculation of minor edges}
\label{edge enhancement fig}
\end{figure}

To further reduce the perceptual difference between the original images and the JND images, we also preserve important edges in the prefiltering process. The reason is that the minor edges in the original images are important details and could easily disappear after JND injection. With edge preservation, we can recover them for better perceptual performance. The flow is as follows: First, we use two Canny edge detectors with different strengths. One is for main edge detection, and the other is for detailed edge detection.
Second, as shown in Fig.~\ref{edge enhancement fig}, we calculate the difference between the main edges and the detailed edges, which represents the minor edges that we want to recover. After finding the minor edges, we replace the pixels of the minor edges in the JND images with the pixels of the same areas in the original images to recover the disappeared pixels due to JND injection. In this way, our JND images can preserve important details and achieve higher perceptual quality.

\subsection{IQA-guided selection}
\subsubsection{Overview}

The desired training data is required to be similar to the original reconstructed image after decoding, which means that we should consider the distortions caused by the codec in the selection of candidates. To accommodate this requirement, we will encode the JND prefiltered images, reconstruct them, and crop them for IQA selection to select the golden data according to the perceptual quality of these reconstructed results, as shown in steps (2) to (4).

Fig.~\ref{groundtruth} illustrates the flow of IQA-guided selection (Step 4). In this process, we will compare the IQA value of the original reconstruction patches with the IQA value of the JND-rec patches, and use this comparison and the additional constraint for smooth areas to select the appropriate JND-rec patches with the largest scale factor $\alpha$ under the quality constraint. The larger scale factor $\alpha$ implies a higher JND value for a lower bitrate. The final training data will be the combination of these selected best JND patches.

\subsubsection{IQA and selection criteria}

The IQA used in this paper is the Neural Image Assessment (NIMA)\cite{NIMA}, which is a no-reference IQA proposed by Google. The reason we chose it is that NIMA, which is trained on the Aesthetic Visual Analysis (AVA) dataset\cite{ava}, can predict image quality with a high correlation with human perception and also works for small patches. In addition, the non-reference IQA is more flexible in training data selection compared to reference IQAs~\cite{9223737} because it does not require the prefiltered JND images to be the same as the original decoded images and can help select golden data perceptually better than the original ones.

The best candidate is selected on the basis of two criteria. The first criterion is The first criterion is the similarity in NIMA rating between the JND-rec patches and the original reconstruction patches. The reason is that the training data should be perceptually lossless compared to the original images after decoding. Thus, we will first evaluate the NIMA rating of different scaled JND-rec patches and the original reconstruction patch. Next, we will find the JND-rec patch with the closest larger rating to the original reconstruction patch and the largest scale factor $\alpha$ as the best for this area. Also, since there could be a small ambiguity in the IQA evaluation, it would be acceptable if the rating of the JND-rec patch was slightly lower than the rating of the original reconstruction patch when the closest larger condition does not meet all available ratings.

The first criterion is applied to all patches except smooth areas, such as human faces. In smooth areas, the JND scales could easily be overestimated because the IQA value could be similar regardless of whether minor but important details disappear or not. A smooth area is a patch with small mean absolute errors (MAE) at $\alpha$ = 0.1. For these areas, we will impose an additional MAE constraint as the second criterion to filter possible candidates before applying the first criterion. This criterion is as follows: First, we check if a patch is smooth or not. If not, apply the first criterion. For a smooth patch, possible patch candidates should have MAE $\leq 1.2\times$MAE at $\alpha$ = 0.1. Only these candidates will be considered for the first criterion. With this, we can set an upper bound on the scaling factor $\alpha$ to avoid JND overestimation in smooth areas and generate training data with better perceptual quality.

\section{IQNet : IQA-guided JND Prefiltering Network}

\subsection{Network architecture}
The JND prefiltering network can be designed to learn filtered images directly or learn the JND values that are the difference between the input images and the training data, which is similar to the super-resolution case. However, the JND value is usually much smaller, which can be learned with a lightweight residual structure to reduce computational complexity for high-definition videos. 

\begin{figure}[tb]
\centering
\includegraphics[height=!,width=0.8\linewidth,keepaspectratio=true]{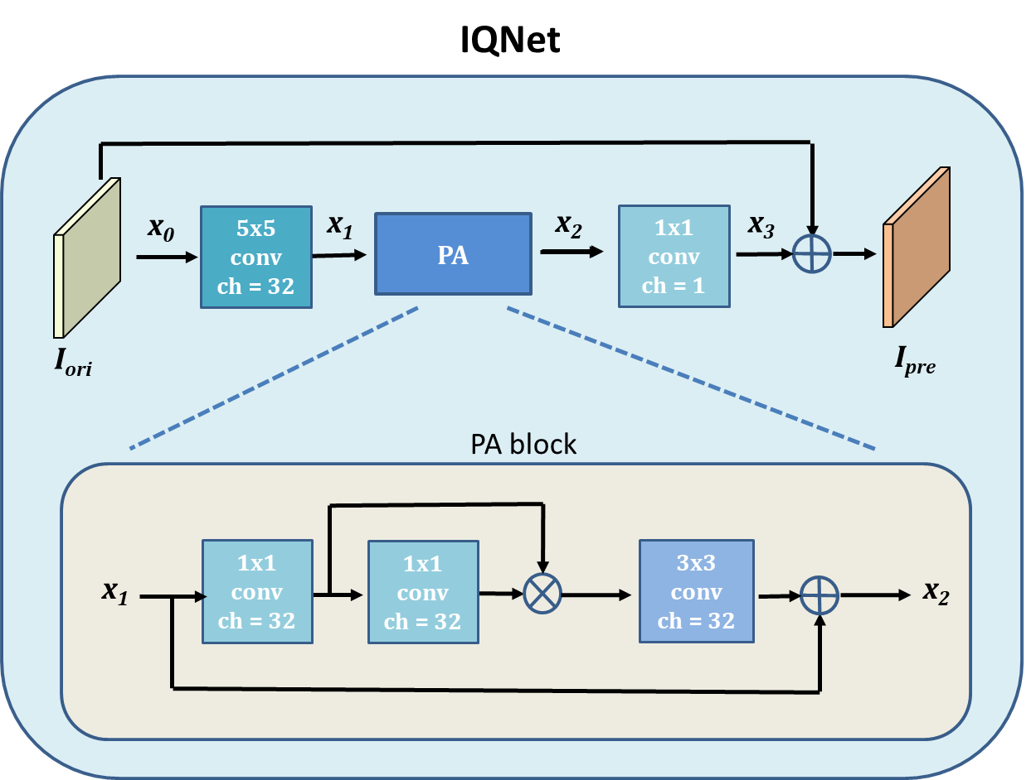}
\caption{The network architecture of the IQNet}
\label{iqnet}
\end{figure}

The proposed network is shown in Fig.~\ref{iqnet}, which is a residual structure with a 5x5 convolution and a pixel attention block~\cite{pixel} for feature extraction. Pixel attention can generate attention coefficients for all pixels on the feature map and only uses a 1×1 convolution layer and a sigmoid function to obtain the attention maps. This not only improves the performance of the model but also leads to fewer parameters and lower computational complexity compared to other attention schemes, such as channel attention. Additionally, because our IQNet attempts to learn the difference between input and training data, we add shortcuts to help model learning. The JND prefiltering will only apply to the luminance part. Thus, the input and output of the model will only be the luminance of the image. For chrominance, it would stay the same throughout the prefiltering process and will be combined with the prefiltered luminance in the end to produce the JND prefiltered image. 

This model is intentionally designed as a lightweight model (3K parameters, 3K MAC (multiplication and accumulation)). In contrast, CNN-JNQD~\cite{ki2018erjnd} needs 38K parameters, which is not acceptable for higher-definition input.

\subsection{Model consideration for different Qps}
The above network is trained once for the QP27 case and applied to other QPs. On the contrary, in HDR-JNDNet~\cite{HDRNET}, Ki et al. generate the training data and train the JND prefiltering network for each QP individually, which is time-consuming and inconvenient for practical use. In this work, from our experiences, we found that the best scale for the same area will be similar in the IQA-guided selection under different QPs, which implies that we could train only one model in the base QP and apply it on different QPs directly. Thus, the problem is which QP should be the base QP. Our candidates for the base QP are 22, 27, 32, and 37. Among them, for low QP such as QP22, the IQA rating of different JND scaled images could sometimes become similar because the quantization effect is minor in reconstructed images, which makes scale selection difficult. For higher QPs such as QP32 and QP37, the IQA evaluation could be inaccurate due to blurry images in high QPs, and the risk of misprediction in selecting the JND scale could be greater. As a result, we choose QP27 as our base QP and train our model referring to the training data provided at QP27.

\section{Experimental Results of Training Data}

\subsection{The results of training data}
To build our dataset, we use 7,824 images from the first frame of the 39K selected video clips in Vimeo-90K~\cite{vimeo} with 448$\times$256 resolution. Vimeo-90K is chosen due to its large variety of scenes and actions to help network training.

\begin{figure}[tb]
\centering
     \begin{subfigure}[b]{0.49\linewidth}
         \centering
         \includegraphics[width=\linewidth]{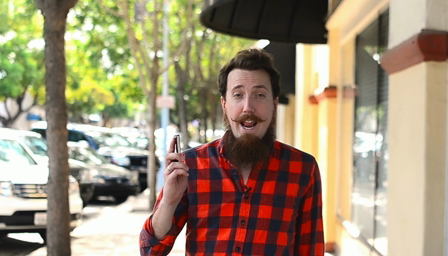}
         \caption{original input}
         \label{ori1717}
     \end{subfigure}
     \begin{subfigure}[b]{0.49\linewidth}
         \centering
         \includegraphics[width=\linewidth]{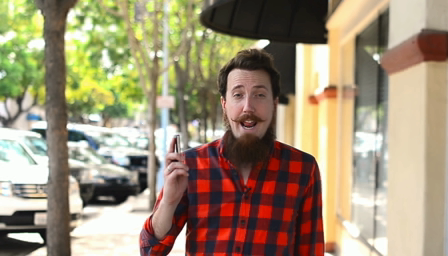}
         \caption{training data}
         \label{gt1717}
     \end{subfigure}
        \caption{Example image before encoding.}
        \label{comparision_ori_gt}
\end{figure}

\begin{figure}[tb]
\centering
     \begin{subfigure}[b]{0.49\linewidth}
         \centering
         \includegraphics[width=\linewidth]{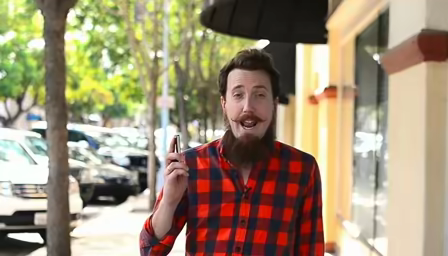}
         \caption{original input}
         \label{ori1717rec27}
     \end{subfigure}
     \begin{subfigure}[b]{0.49\linewidth}
         \centering
         \includegraphics[width=\linewidth]{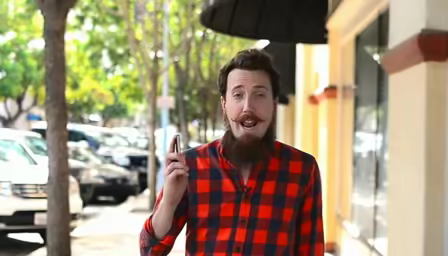}
         \caption{training data}
         \label{gt1717rec27}
     \end{subfigure}
        \caption{The reconstruction images after encoded at QP27.}
        \label{comparision_ori_gt_rec27}
\end{figure}

\begin{figure}[tb]
     \begin{subfigure}[b]{0.49\linewidth}
         \includegraphics[width=\linewidth]{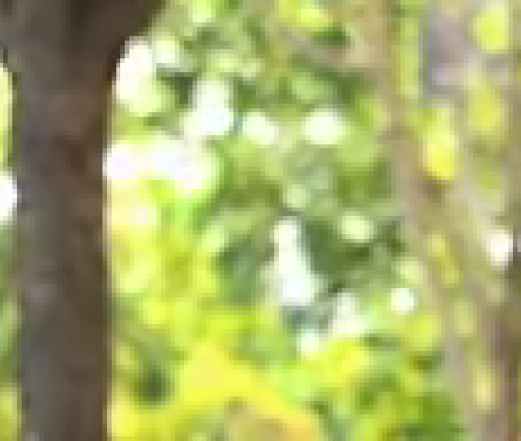}
         \label{bush_ori_rec27}
     \end{subfigure}
     \begin{subfigure}[b]{0.49\linewidth}
         \includegraphics[width=\linewidth]{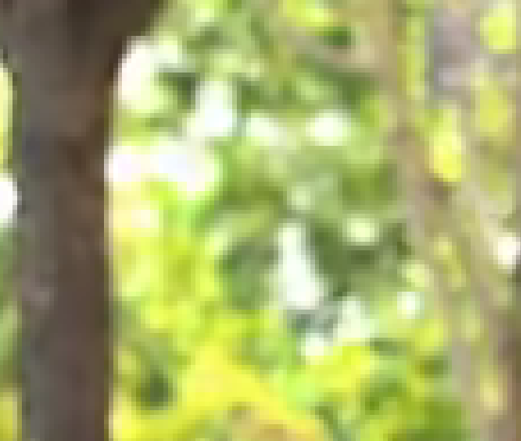}
         \label{bush_gt_rec27}
     \end{subfigure}
     \begin{subfigure}[b]{1\linewidth}
         \centering
         \includegraphics[width=\linewidth]{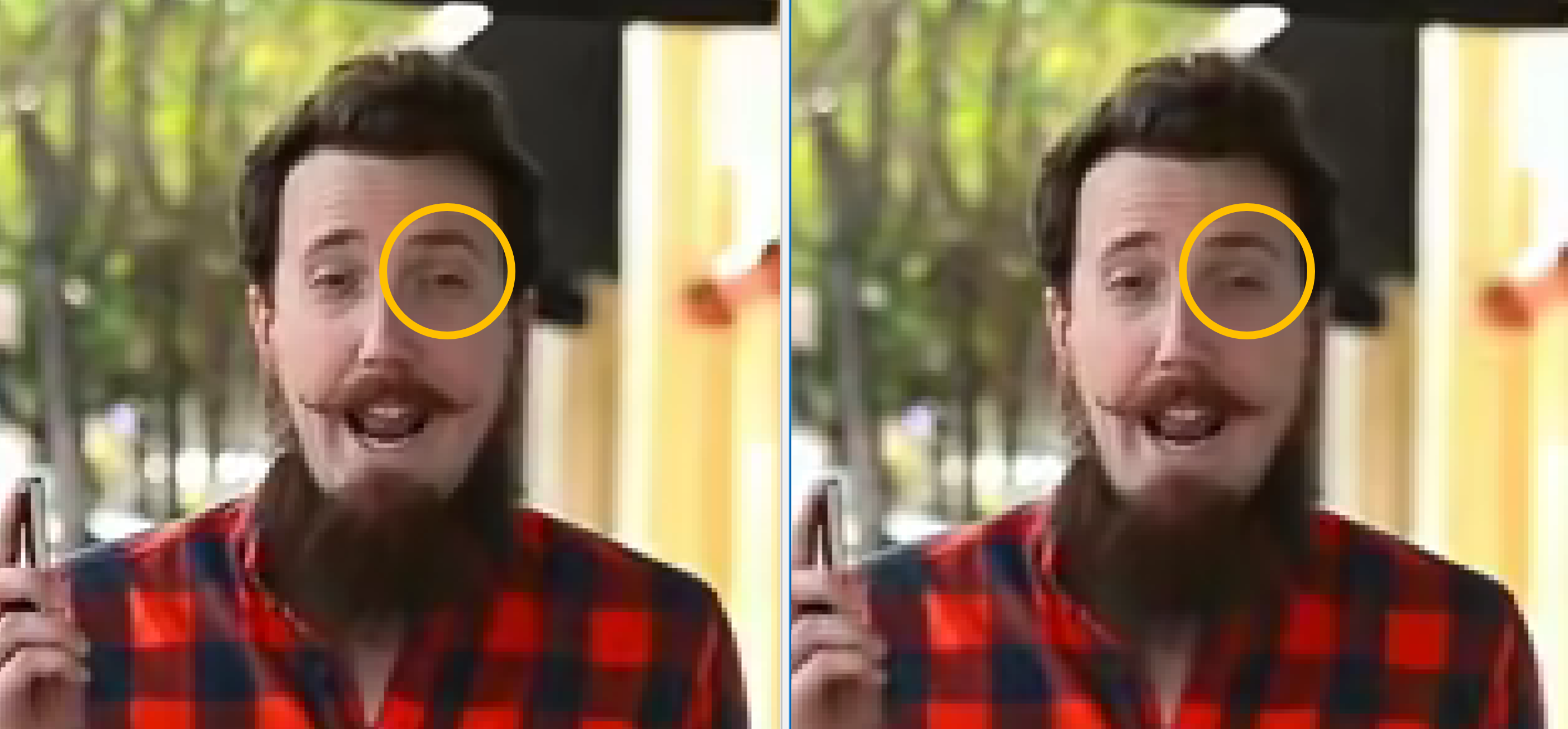}
         \label{eye_gt_rec27}
     \end{subfigure} 
        \caption{The enlarged reconstruction images of the original input (left) and the training data (right) encoded at QP27. }
        \label{comparision_bush}
\end{figure}

Fig.~\ref{comparision_ori_gt} and Fig.~\ref{comparision_ori_gt_rec27} show visual comparisons between the original input image and our training data and their reconstructed versions encoded with VVC in QP27. From comparisons, we can find that the perceptual quality of the original input and our data are similar, regardless of whether we encode them or not. However, this JND prefiltered image requires around 10\% lower bit rate than the original because some unperceivable details of the original image are removed after JND prefiltering without loss of perceptual quality. As shown in Fig.~\ref{comparision_bush}, we can see that the bush in this area loses details and becomes smoother after JND prefiltering; however, removing this information is unperceivable due to the low sensitivity to texture regions such as bush in the human visual system, and therefore we could successfully improve bitrate without a drop in perceptual quality. A similar situation also occurs in the right eye, as shown in Fig.~\ref{comparision_bush}.

\subsection{Ablation studies of the proposed training data generation}
\begin{figure}[tb]
    \centering
     \begin{subfigure}[b]{1\linewidth}
     \centering
     \includegraphics[width=\linewidth]{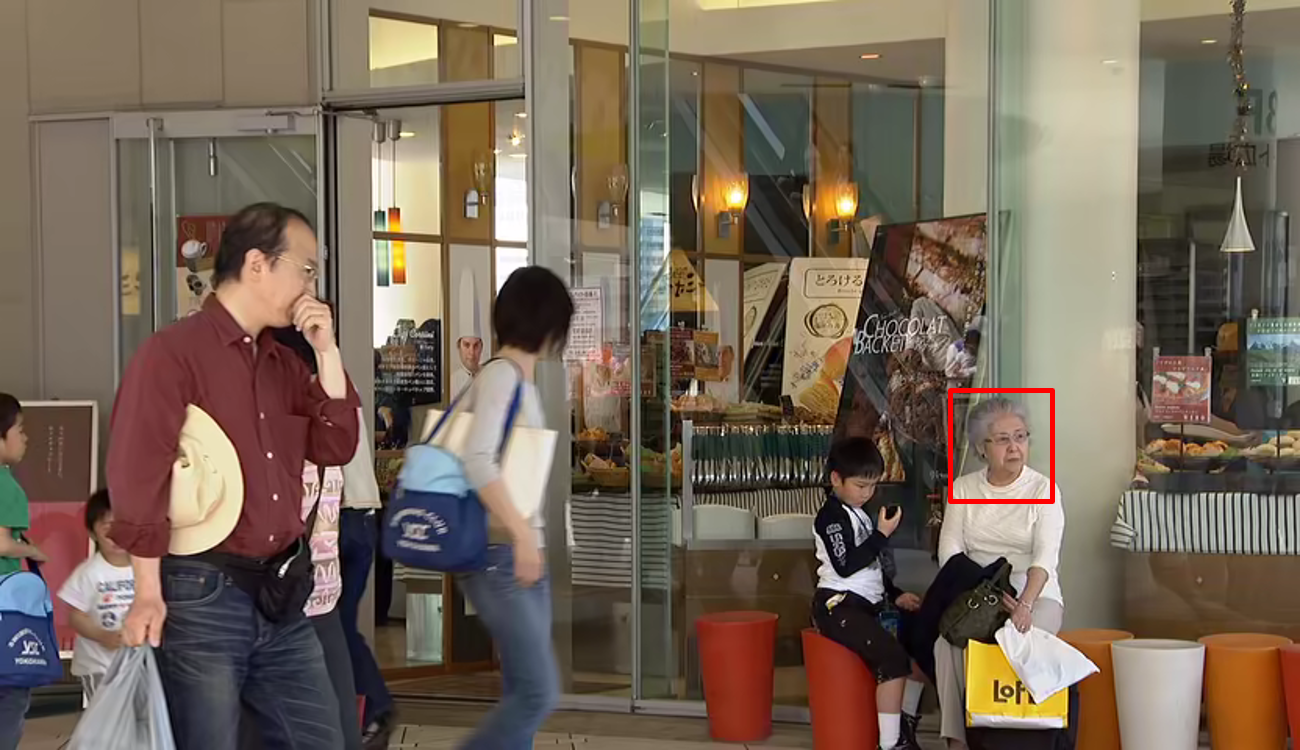}
     \caption{our JND}
     \label{bqmallspot1}
     \end{subfigure}
     \begin{subfigure}[b]{0.49\linewidth}
         \centering
         \includegraphics[width=\linewidth]{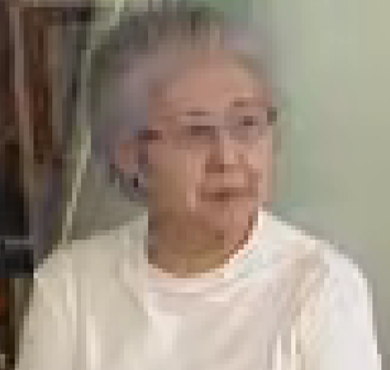}
         \caption{ERJND}
         \label{erjnd_face}
     \end{subfigure}
     \begin{subfigure}[b]{0.49\linewidth}
         \centering
         \includegraphics[width=\linewidth]{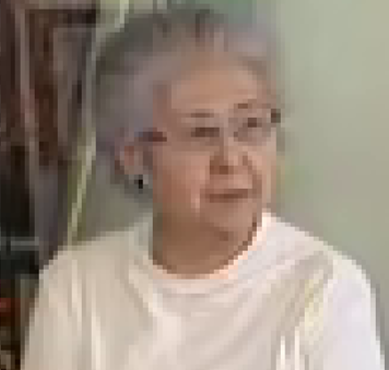}
         \caption{our JND}
         \label{gt_face}
     \end{subfigure}
        \caption{The reconstruction image from \textit{BQMall} encoded at QP27.}
        \label{prefilter_result}
\end{figure}

\textbf{Comparison with ERJND} 
To show the difference in results between ERJND and our JND scheme, we used the first frame of $BQMall$ as a test image and generated the reconstructed results of this image with two different prefiltering methods under the same JND scale in QP27, as shown in Fig.~\ref{prefilter_result}. We could observe that the reconstructed image after ERJND prefiltering (Fig.~\ref{prefilter_result}(b)) is blurrier than ours (Fig.~\ref{prefilter_result}(c)). In addition, that image has obvious artifacts and a blocking effect. On the contrary, ours is smoother and avoids those problems.

\begin{figure}[tb]
    \centering
     \begin{subfigure}[b]{0.3\linewidth}
         \centering
         \includegraphics[width=\linewidth]{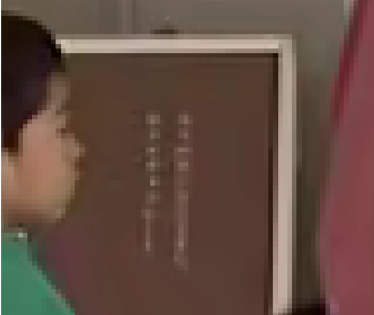}
         \caption{original input}
         \label{edge1}
     \end{subfigure}
     \hfill
     \begin{subfigure}[b]{0.3\linewidth}
         \centering
         \includegraphics[width=\linewidth]{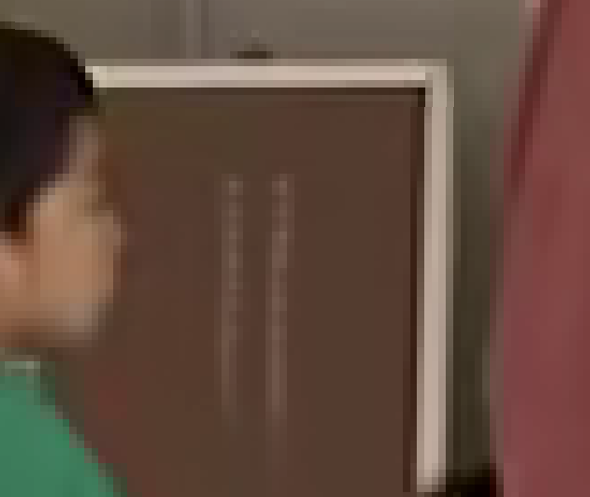}
         \caption{without}
         \label{edge2}
     \end{subfigure}     
     \hfill
     \begin{subfigure}[b]{0.3\linewidth}
         \centering
         \includegraphics[width=\linewidth]{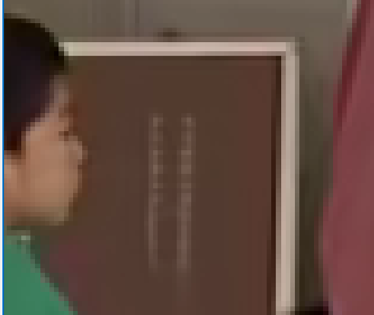}
         \caption{with}
         \label{edge3}
     \end{subfigure}
        \caption{The reconstruction images for edge preservation test, when encoded at QP27. (a) is for original input. (b) and (c) are training data without/with edge preservation, respectively.}  
        \label{edge_result3}
\end{figure}

\textbf{Effect of Edge Preservation} Fig.~\ref{edge_result3} illustrates the comparison to show the effect of edge preservation. We can observe that without this, the reconstructed JND prefiltered image (Fig.~\ref{edge_result3}(b)) could be blurry due to the disappearance of minor edges. However, with edge preservation, important details, such as the boy's face, are recovered, and its reconstructed image (Fig.~\ref{edge_result3}(c)) is more similar to the original reconstructed image (Fig.~\ref{edge_result3}(a)).

\begin{figure}
    \centering
     \begin{subfigure}[b]{0.3\linewidth}
         \centering
         \includegraphics[width=\linewidth]{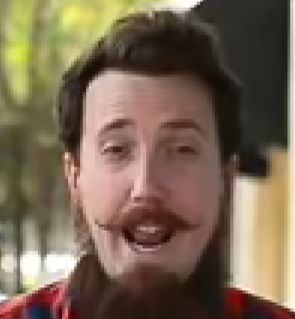}
         \caption{original input}
         \label{head1}
     \end{subfigure}
     \hfill
     \begin{subfigure}[b]{0.3\linewidth}
         \centering
         \includegraphics[width=\linewidth]{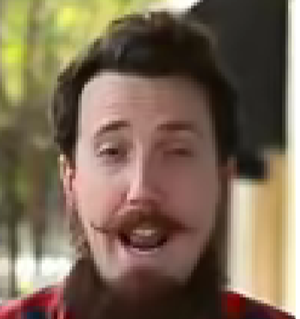}
         \caption{without}
         \label{head2}
     \end{subfigure}     
     \hfill
     \begin{subfigure}[b]{0.3\linewidth}
         \centering
         \includegraphics[width=\linewidth]{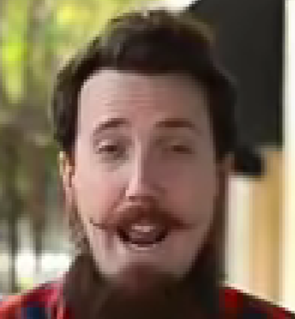}
         \caption{with}
         \label{head3}
     \end{subfigure}
        \caption{The reconstructed images for the MAE constraint test, when encoded at QP27, are as follows: (a) is for the original input. (b) and (c) are training data without/with the MAE constraint, respectively.}
        \label{head_result3}
\end{figure}

\textbf{Effect of the MAE Constraint on Smooth Areas}
In our IQA-guided selection for training data generation, we set restrictions not only on the IQA rating but also on the MAE for smooth areas. As Fig.~\ref{head_result3} shows, if we did not set the additional constraint for smooth areas such as the human face, some important features such as the right eye would lose detailed information and become blurry. But with this constraint, we could retain the critical details and achieve better perceptual quality.

\begin{figure}[tb]
     \centering
     \begin{subfigure}[b]{0.8\linewidth}
     \centering
     \includegraphics[width=\linewidth]{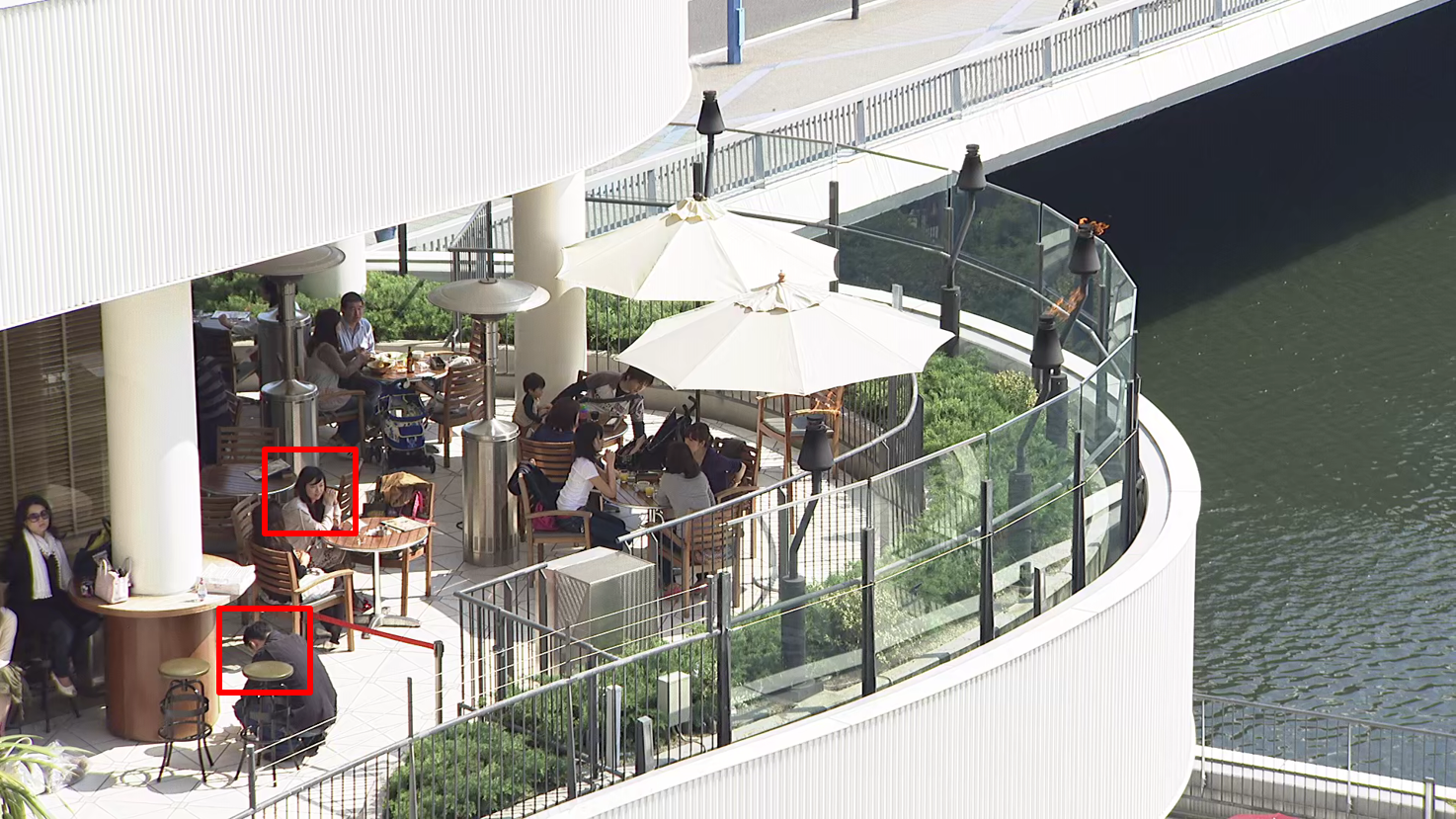}
     \caption{anchor}
     \label{terrscespot1}
     \end{subfigure}
     \begin{subfigure}[b]{0.4\linewidth}
         \centering
         \includegraphics[width=\linewidth]{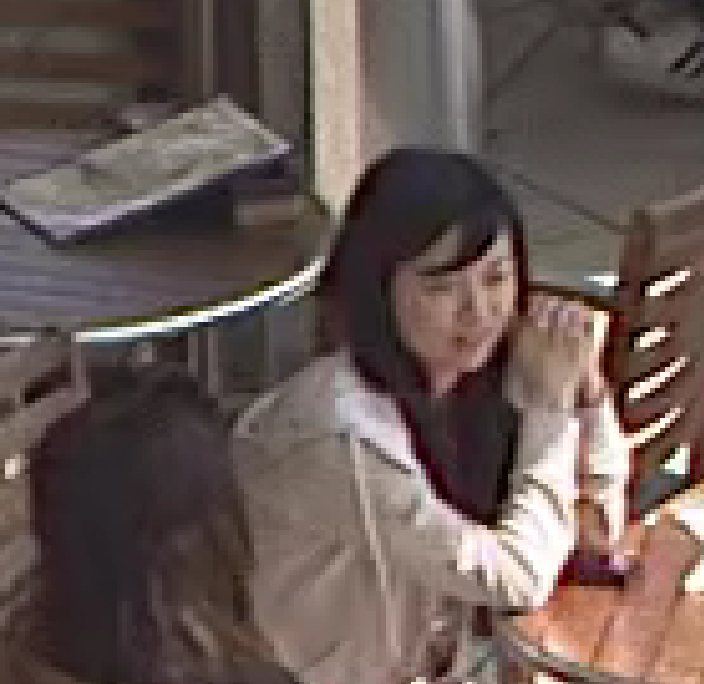}
         \caption{anchor}
         \label{face_ori_rec27}
     \end{subfigure}
     \begin{subfigure}[b]{0.4\linewidth}
         \centering
         \includegraphics[width=\linewidth]{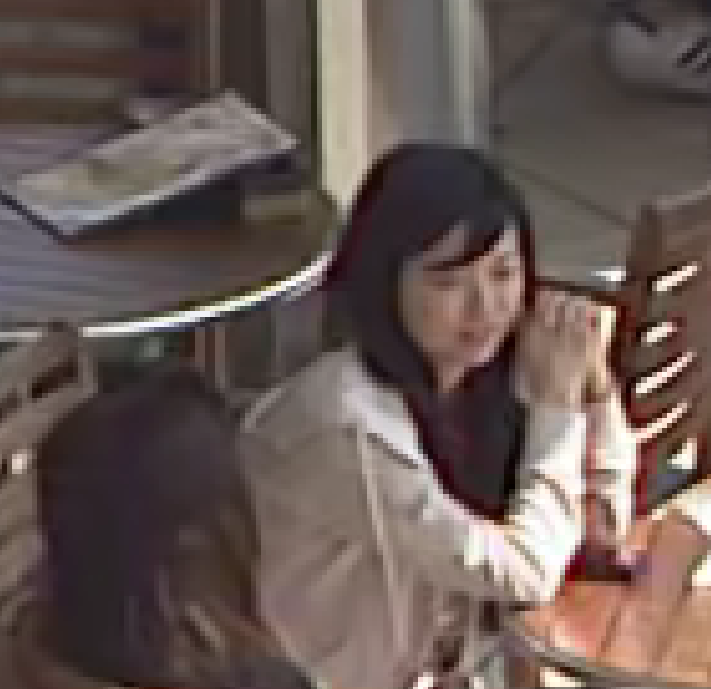}
         \caption{IQNet}
         \label{face_iq_rec27}
     \end{subfigure}
     \hfill
     \begin{subfigure}[b]{0.4\linewidth}
         \centering
         \includegraphics[width=\linewidth]{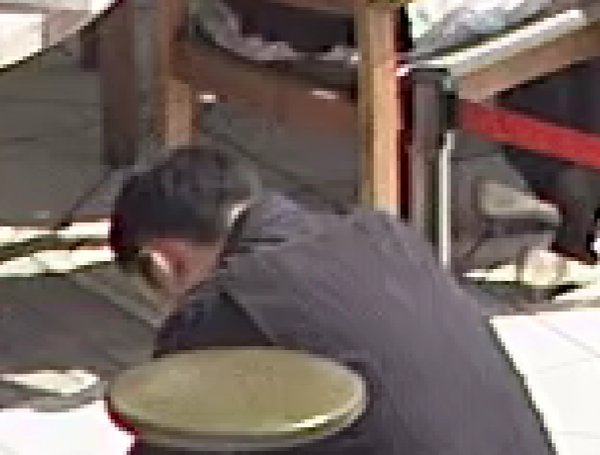}
         \caption{anchor}
         \label{man_ori_rec27}
     \end{subfigure}
     \begin{subfigure}[b]{0.4\linewidth}
         \centering
         \includegraphics[width=\linewidth]{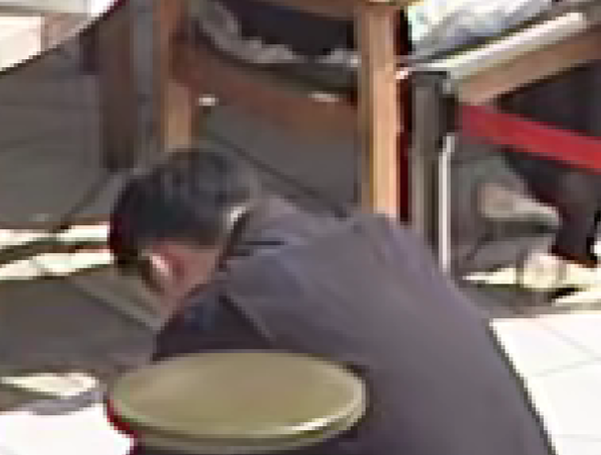}
         \caption{IQNet}
         \label{man_iq_rec27}
     \end{subfigure}     
        \caption{The reconstructed first frame in $BQterrace$ encoded at QP27. (PSNR, bitrate) for the anchor and IQNet are (38.16, 1030030) and (35.38, 78247), respectively.}
        \label{comparision_face}
\end{figure}

\begin{figure}[tb]
    \centering
     \begin{subfigure}[b]{0.8\linewidth}
     \centering
     \includegraphics[width=\linewidth]{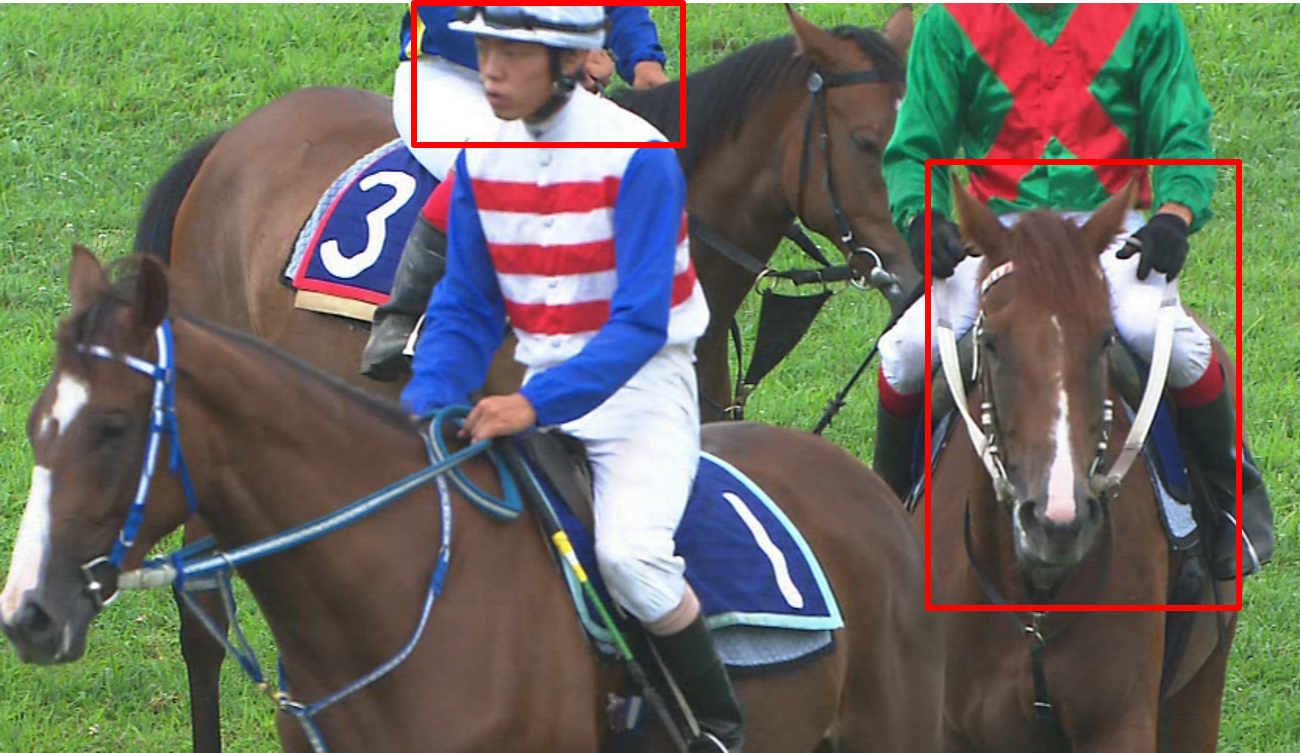}
     \caption{anchor}
     \label{horsespot32}
     \end{subfigure}    
     \begin{subfigure}[b]{0.3\linewidth}
         \centering
         \includegraphics[width=\linewidth]{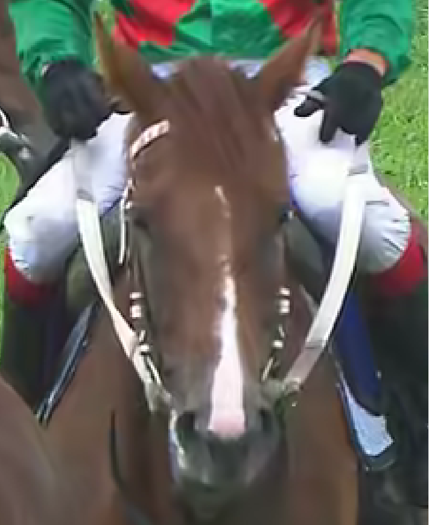}
         \caption{anchor}
         \label{horse32_1}
     \end{subfigure}
     \begin{subfigure}[b]{0.3\linewidth}
         \centering
         \includegraphics[width=\linewidth]{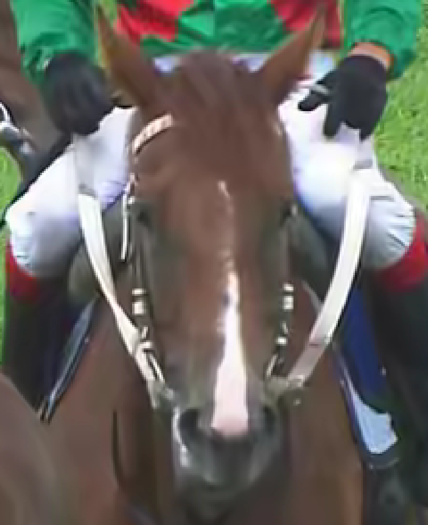}
         \caption{IQNet}
         \label{horse32_2}
     \end{subfigure}    

     \begin{subfigure}[b]{0.3\linewidth}
         \centering
         \includegraphics[width=\linewidth]{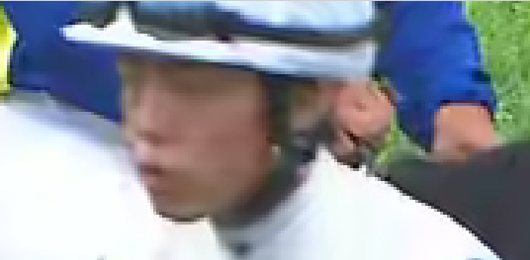}
         \caption{anchor}
         \label{horse32_3}
     \end{subfigure}
     \begin{subfigure}[b]{0.3\linewidth}
         \centering
         \includegraphics[width=\linewidth]{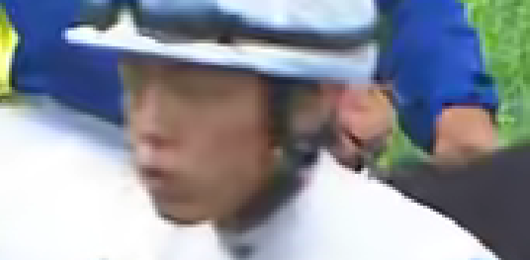}
         \caption{IQNet}
         \label{horse32_4}
     \end{subfigure} 
        \caption{The reconstructed first frame in $RaceHorses$ encoded at QP32. (PSNR, bitrate) for the anchor and IQNet are (34.52, 1956) and 32.79, 1667), respectively.}
        \label{horse_result32}
\end{figure}

\section{Experimental Results of IQNet}
\subsection{Implementation details of IQNet}

IQNet is implemented with PyTorch and trained on Nvidia 2080Ti GPUs. The model uses only the luminance of the images and will combine it with the original chrominance to produce the final JND prefiltered image. For each training batch, we randomly extract a 64x64 patch as input with the batch size set to 128. We train our model for 2,500 epochs with the Adam optimizer, setting the learning rate to 0.001 with the rate decayed by half every 250 epochs. We used 18 HEVC test sequences listed in Table~\ref{HEVC seq} as a test set and encoded them using VVC VTM-11.0 for bitrate, PSNR, and image quality evaluation. For the following comparison, we use the VVC reconstructed video without any prefiltering as our anchor.

\begin{table}[tb]
\centering
\caption{List of 18 HEVC test sequences as our test set.}
\label{HEVC seq}
\begin{tabular}{|l|l|l|}
\hline
Class & Resolution  & Sequence                                                                                            \\ \hline
A     & 2560 $\times$ 1600 & PeopleOnStreet, Traffic                                                                             \\ \hline
B     & 1920 $\times$ 1080 & \begin{tabular}[c]{@{}l@{}}BQTerrace,   BasketballDrive. Cactus, \\ Kimono1, ParkScene\end{tabular} \\ \hline
C     & 832 $\times$ 480   & \begin{tabular}[c]{@{}l@{}}BQMall,   BasketballDrill, PartyScene,\\ RaceHorses\end{tabular}         \\ \hline
D     & 416 $\times$ 240   & \begin{tabular}[c]{@{}l@{}}BQSquare,   BasketballPass, \\ BlowingBubbles, RaceHorses\end{tabular}   \\ \hline
E     & 1280 $\times$ 720  & FourPeople,   Johnny, KristenAndSara                                                                \\ \hline
\end{tabular}

\end{table}

\begin{figure*}[tb]
\centering
\includegraphics[height=!,width=0.49\linewidth,keepaspectratio=true]{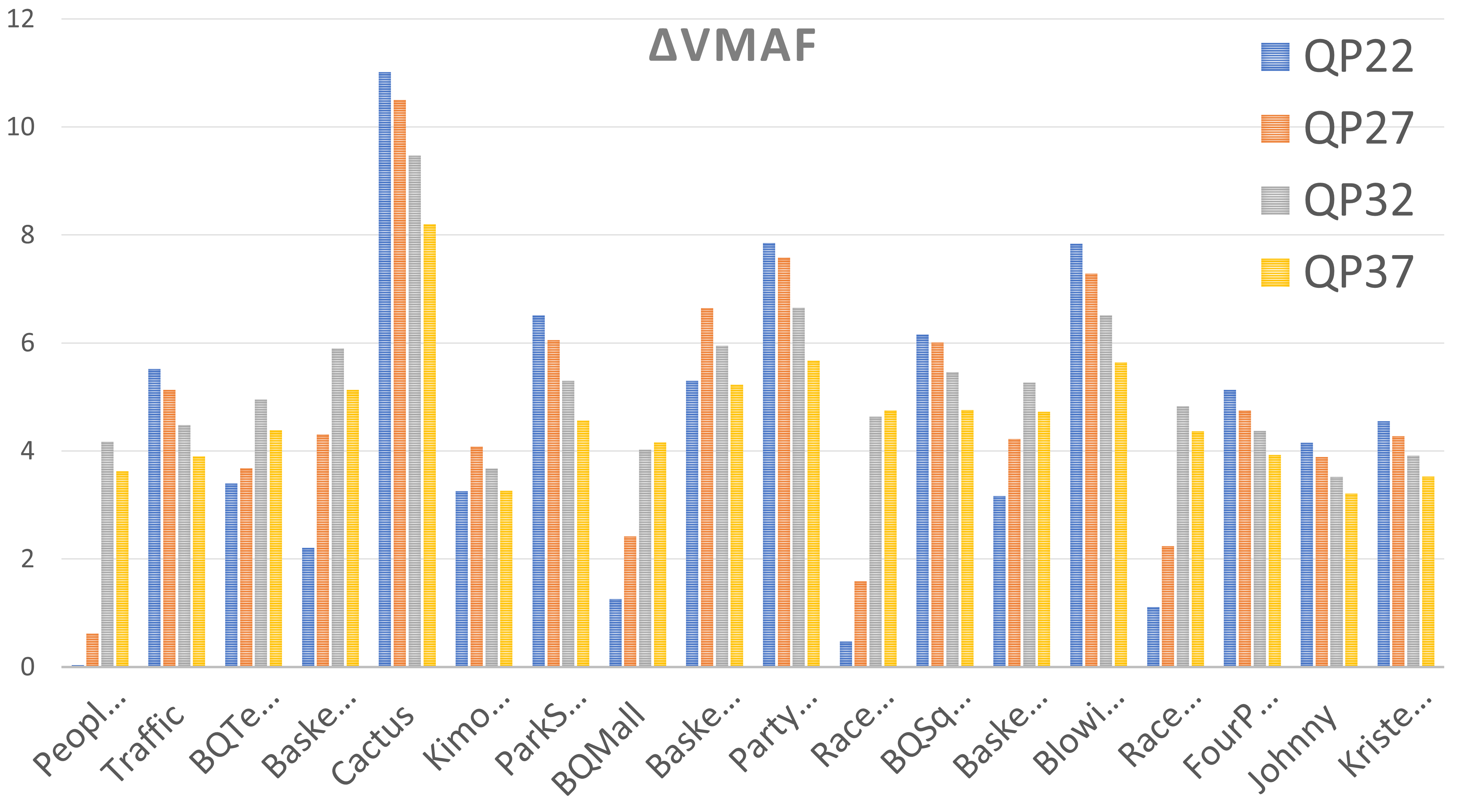}
\includegraphics[height=!,width=0.49\linewidth,keepaspectratio=true]{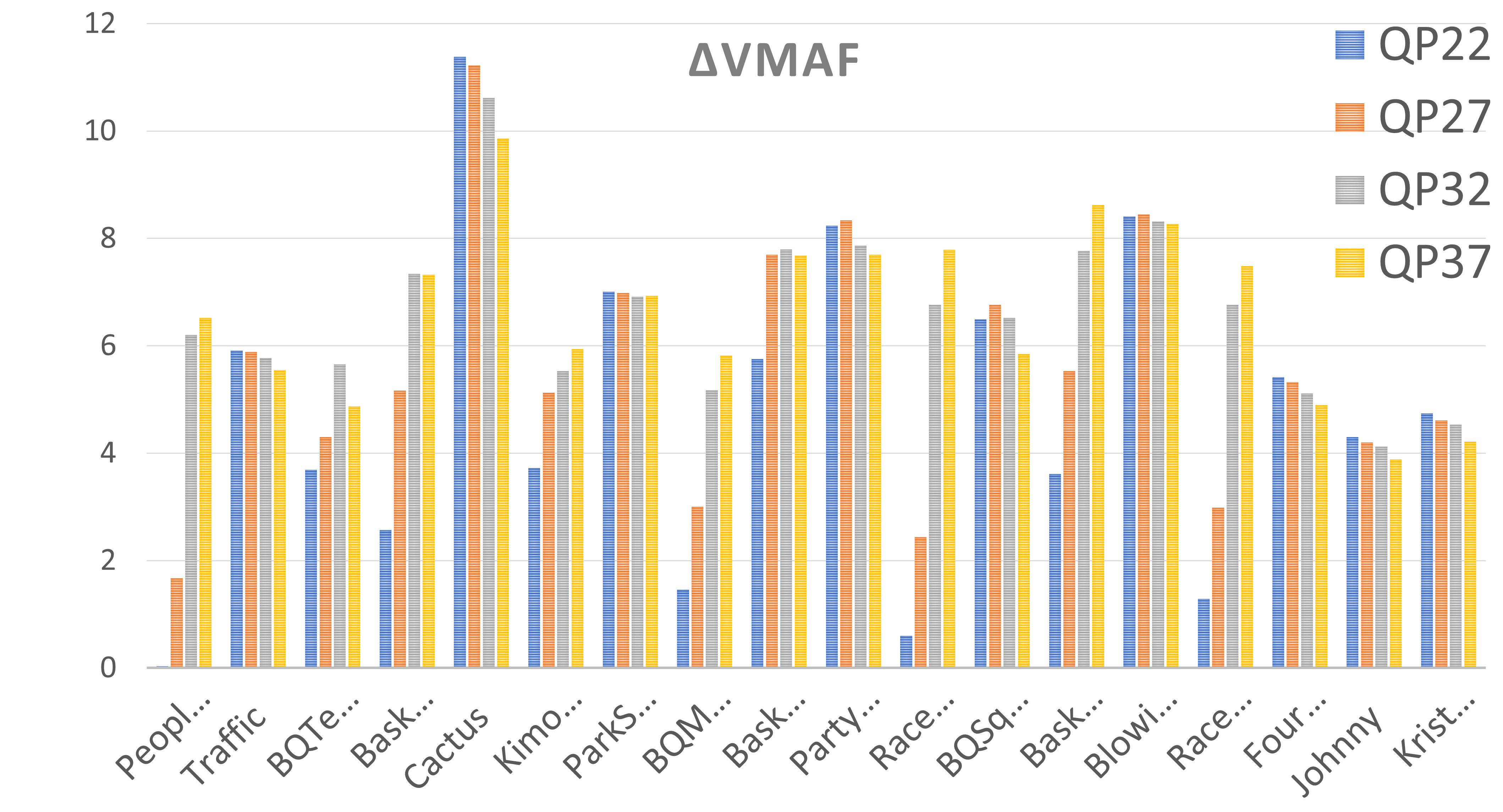}
\caption{VMAF drop (anchor - IQNet) of the reconstructed videos with IQNet compared to the anchor. Left: all intra profile. Right: low delay P profile.}
\label{vmaf_result}
\end{figure*}

\begin{figure*}[tb]
\centering
\includegraphics[height=!,width=0.49\linewidth,keepaspectratio=true]{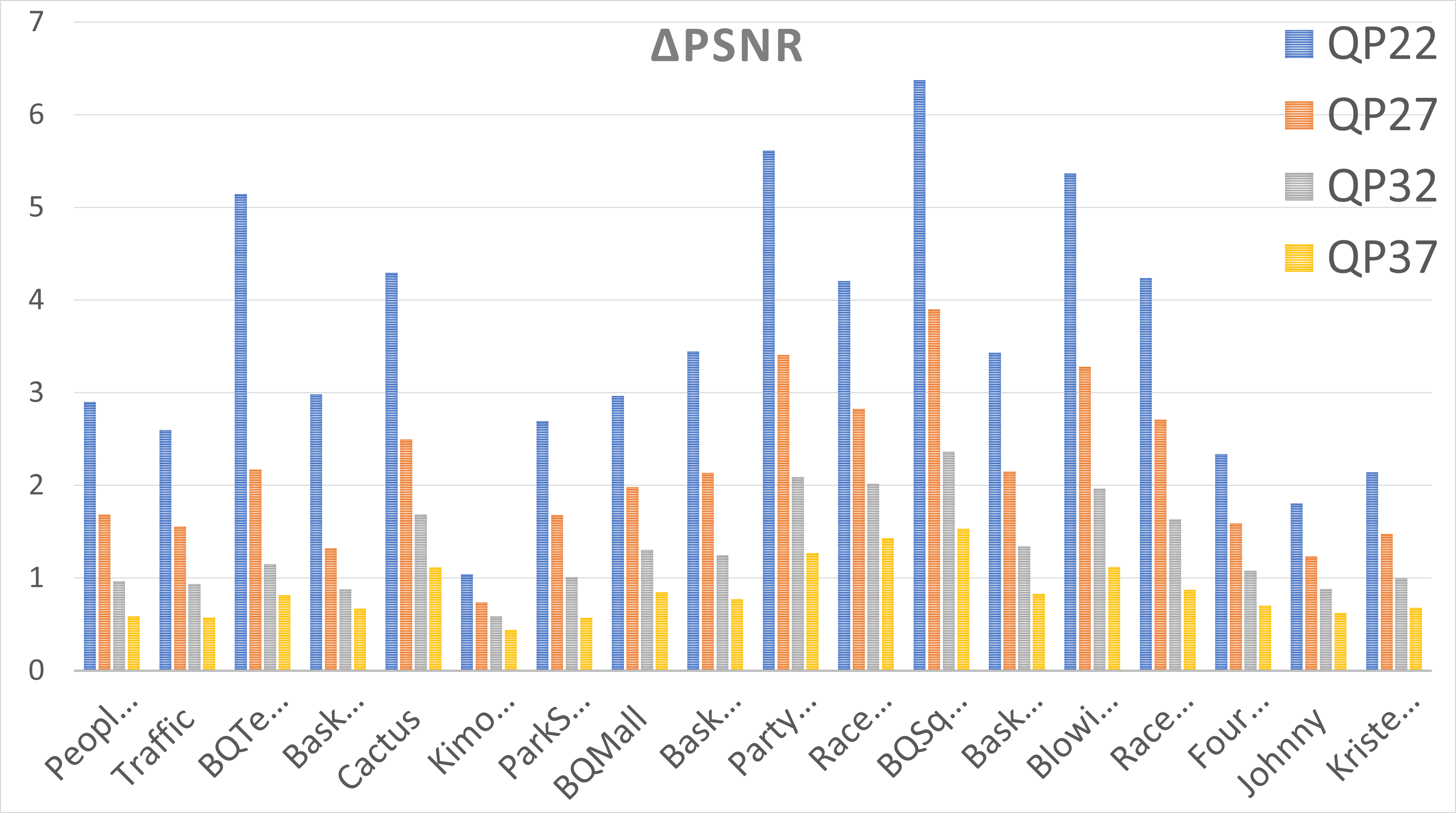}
\includegraphics[height=!,width=0.49\linewidth,keepaspectratio=true]{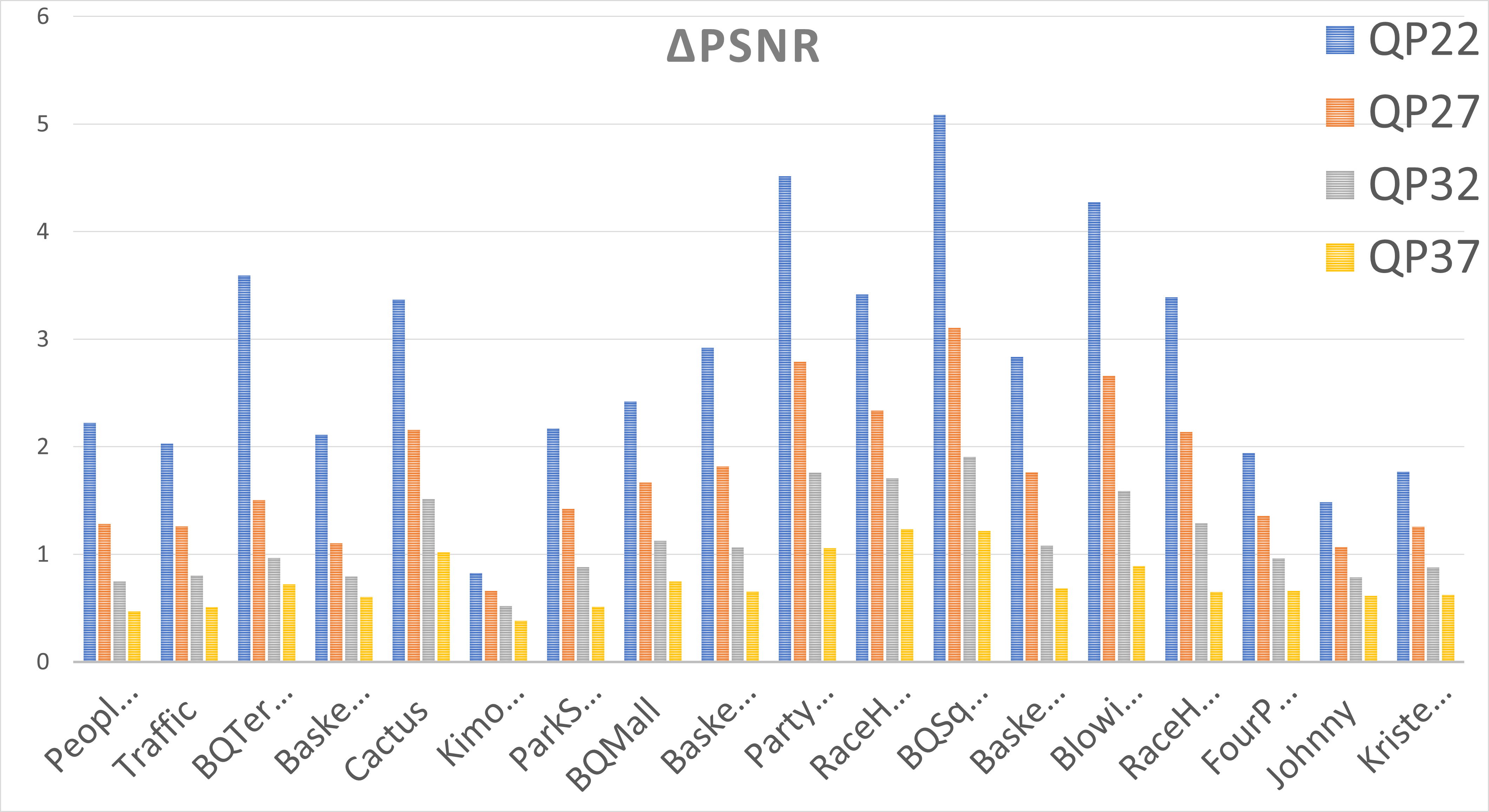}
\caption{PSNR drop (anchor - IQNet) of the reconstructed videos with IQNet compared to the anchor. Left: all intra profile. Right: low delay P profile.}
\label{psnr_result}
\end{figure*}

\begin{figure*}[tb]
\centering
\includegraphics[height=!,width=0.49\linewidth,keepaspectratio=true]{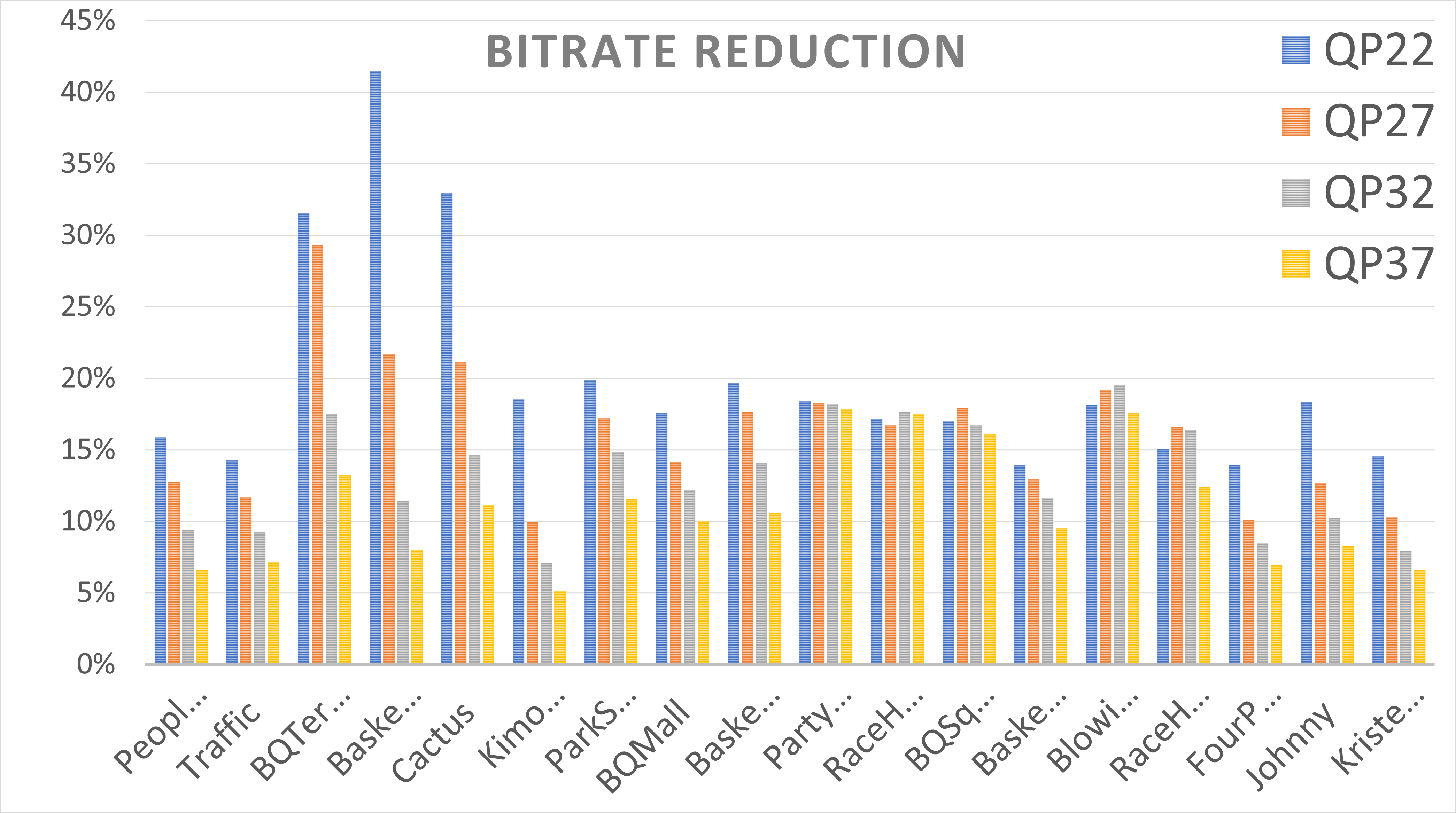}
\includegraphics[height=!,width=0.49\linewidth,keepaspectratio=true]{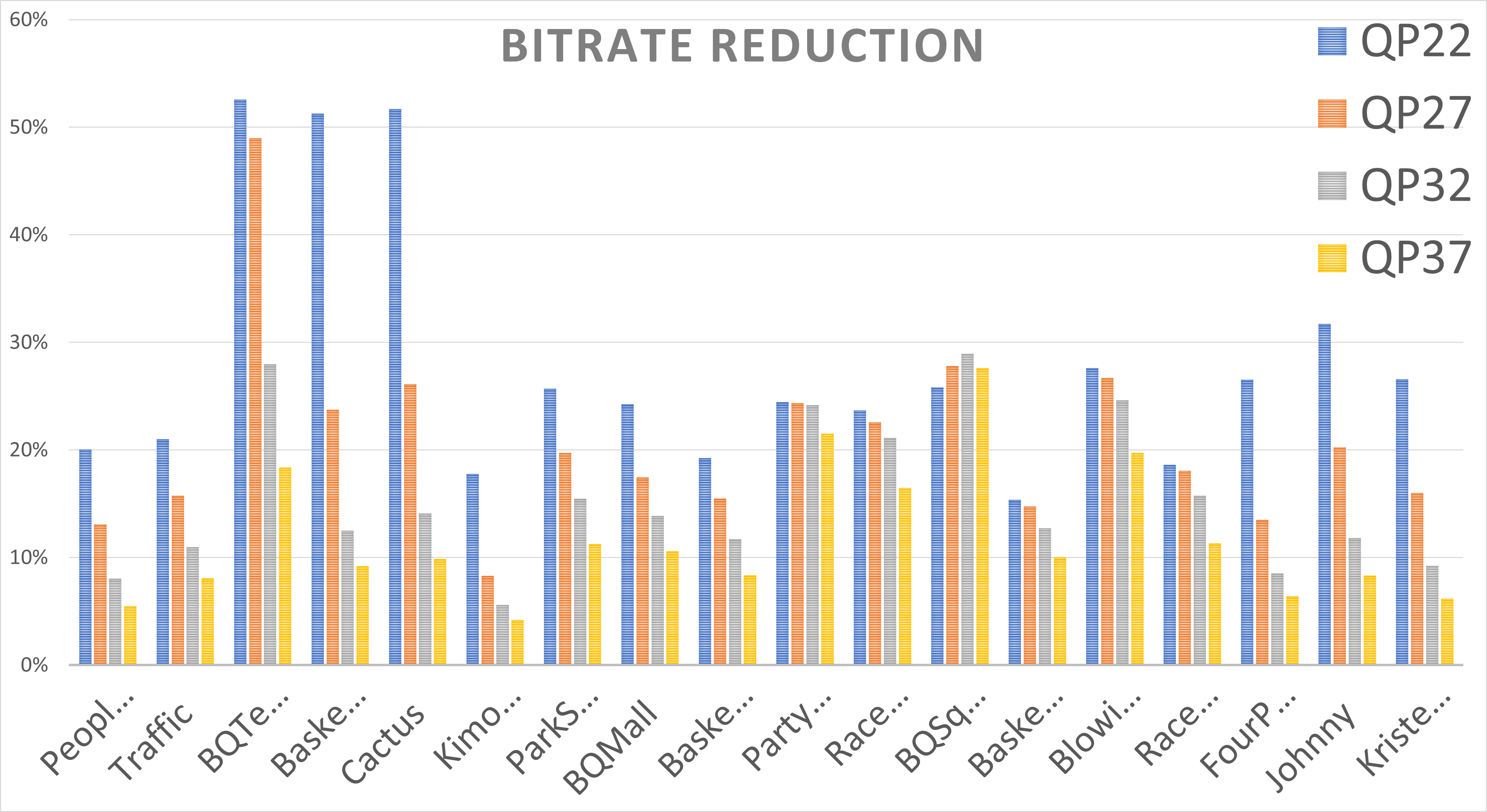}
\caption{Bitrate reduction ((anchor-IQNet)/anchor) of the reconstructed videos with IQNet compared to the anchor. Left: all intra profile. Right: low delay P profile.}
\label{bitrate_result}
\end{figure*}

\begin{table}
\centering
\caption{The bitrate and VMAF rating for all intra-coded training images with the proposed IQNet and the method of training data generation, where (T/I) stands for VMAF of the prefiltered video before encoding for training data and IQNet, respectively and VMAF$_r$ stands for VMAF of the reconstructed video.}
\label{table_gt}
\begin{tabular}{|c|c|cc|cc|}
\hline
\multirow{2}{*}{\begin{tabular}[c]{@{}c@{}}Image\\ (T/I)\end{tabular}}          & \multirow{2}{*}{QP} & \multicolumn{2}{c|}{training data}      & \multicolumn{2}{c|}{IQNet}              \\ \cline{3-6} 
                                                                                &                     & \multicolumn{1}{c|}{Bitrate} & VMAF$_r$ & \multicolumn{1}{c|}{Bitrate} & VMAF$_r$ \\ \hline
\multirow{4}{*}{\begin{tabular}[c]{@{}c@{}}image1\\ (92.42/91.58)\end{tabular}} & 22                  & \multicolumn{1}{c|}{4643.04} & 90.73    & \multicolumn{1}{c|}{4676.16} & 89.94    \\ \cline{2-6} 
                                                                                & 27                  & \multicolumn{1}{c|}{2901.12} & 87.71    & \multicolumn{1}{c|}{2927.52} & 87.17    \\ \cline{2-6} 
                                                                                & 32                  & \multicolumn{1}{c|}{1801.44} & 82.83    & \multicolumn{1}{c|}{1821.12} & 81.39    \\ \cline{2-6} 
                                                                                & 37                  & \multicolumn{1}{c|}{1116.48} & 72.66    & \multicolumn{1}{c|}{1108.32} & 72.54    \\ \hline
\multirow{4}{*}{\begin{tabular}[c]{@{}c@{}}image2\\ (88.63/90.24)\end{tabular}} & 22                  & \multicolumn{1}{c|}{8698.08} & 87.10    & \multicolumn{1}{c|}{8849.76} & 88.48    \\ \cline{2-6} 
                                                                                & 27                  & \multicolumn{1}{c|}{5430.24} & 84.61    & \multicolumn{1}{c|}{5597.76} & 85.62    \\ \cline{2-6} 
                                                                                & 32                  & \multicolumn{1}{c|}{3300.96} & 78.37    & \multicolumn{1}{c|}{3418.08} & 79.52    \\ \cline{2-6} 
                                                                                & 37                  & \multicolumn{1}{c|}{1976.16} & 67.38    & \multicolumn{1}{c|}{2018.88} & 66.55    \\ \hline
\end{tabular}

\end{table}

\begin{table*}[tb]
\centering
\caption{VMAF rating and bitrate comparisons among VVC anchor, IQNet, and CNN-JNQD~\cite{ki2018erjnd} prefiltering. VMAF$_p$ stands for VMAF of the prefiltered video before encoding to show the effect of the prefiltering. VMAF$_r$ stands for VMAF of the reconstructed video. VMAF is calculated relative to the original input. The $\Delta$ bitrate represents bitrate saving relative to the anchor.}

\label{table_iqcnn3}
\begin{tabular}{|lllllllllll|}
\hline
\multicolumn{11}{|l|}{All intra configuration}                                                                                                                                                                                                                                                                                                                                                                         \\ \hline
\multicolumn{1}{|l|}{\multirow{2}{*}{Sequence}}   & \multicolumn{1}{l|}{\multirow{2}{*}{Resolution}}         & \multicolumn{1}{l|}{\multirow{2}{*}{QP}} & \multicolumn{2}{l|}{Anchor}                                 & \multicolumn{3}{l|}{CNN-JNQD~\cite{ki2018erjnd}}                                                                            & \multicolumn{3}{l|}{IQNet}                                                        \\ \cline{4-11} 
\multicolumn{1}{|l|}{}                            & \multicolumn{1}{l|}{}                                    & \multicolumn{1}{l|}{}                    & \multicolumn{1}{l|}{VMAF$_r$} & \multicolumn{1}{l|}{bitrate} & \multicolumn{1}{l|}{VMAF$_p$} & \multicolumn{1}{l|}{VMAF$_r$} & \multicolumn{1}{l|}{$\Delta$ Bitrate} & \multicolumn{1}{l|}{VMAF$_p$} & \multicolumn{1}{l|}{VMAF$_r$} & $\Delta$Bitrate \\ \hline
\multicolumn{1}{|l|}{\multirow{4}{*}{PartyScene}} & \multicolumn{1}{l|}{\multirow{4}{*}{832 $\times$ 480}}   & \multicolumn{1}{l|}{22}                  & \multicolumn{1}{l|}{99.11}     & \multicolumn{1}{l|}{44648}   & \multicolumn{1}{l|}{92.03}      & \multicolumn{1}{l|}{81.89}     & \multicolumn{1}{l|}{19\%}             & \multicolumn{1}{l|}{92.77}     & \multicolumn{1}{l|}{91.26}     & 18\%            \\ \cline{3-11} 
\multicolumn{1}{|l|}{}                            & \multicolumn{1}{l|}{}                                    & \multicolumn{1}{l|}{27}                  & \multicolumn{1}{l|}{96.25}     & \multicolumn{1}{l|}{27716}   & \multicolumn{1}{l|}{86.65}      & \multicolumn{1}{l|}{80.68}     & \multicolumn{1}{l|}{19\%}             & \multicolumn{1}{l|}{92.77}     & \multicolumn{1}{l|}{88.67}     & 18\%            \\ \cline{3-11} 
\multicolumn{1}{|l|}{}                            & \multicolumn{1}{l|}{}                                    & \multicolumn{1}{l|}{32}                  & \multicolumn{1}{l|}{89.01}     & \multicolumn{1}{l|}{16417}   & \multicolumn{1}{l|}{84.13}      & \multicolumn{1}{l|}{77.25}     & \multicolumn{1}{l|}{15\%}             & \multicolumn{1}{l|}{92.77}     & \multicolumn{1}{l|}{82.36}     & 18\%            \\ \cline{3-11} 
\multicolumn{1}{|l|}{}                            & \multicolumn{1}{l|}{}                                    & \multicolumn{1}{l|}{37}                  & \multicolumn{1}{l|}{75.69}     & \multicolumn{1}{l|}{9168}    & \multicolumn{1}{l|}{83.17}      & \multicolumn{1}{l|}{69.47}     & \multicolumn{1}{l|}{4\%}              & \multicolumn{1}{l|}{92.77}     & \multicolumn{1}{l|}{70.01}     & 18\%            \\ \hline
\multicolumn{1}{|l|}{\multirow{4}{*}{BQTerrace}}  & \multicolumn{1}{l|}{\multirow{4}{*}{1920 $\times$1080}}  & \multicolumn{1}{l|}{22}                  & \multicolumn{1}{l|}{99.41}     & \multicolumn{1}{l|}{181049}  & \multicolumn{1}{l|}{95.82}      & \multicolumn{1}{l|}{89.79}     & \multicolumn{1}{l|}{43\%}             & \multicolumn{1}{l|}{96.61}     & \multicolumn{1}{l|}{96.02}     & 32\%            \\ \cline{3-11} 
\multicolumn{1}{|l|}{}                            & \multicolumn{1}{l|}{}                                    & \multicolumn{1}{l|}{27}                  & \multicolumn{1}{l|}{98.39}     & \multicolumn{1}{l|}{89899}   & \multicolumn{1}{l|}{92.27}      & \multicolumn{1}{l|}{87.41}     & \multicolumn{1}{l|}{37\%}             & \multicolumn{1}{l|}{96.61}     & \multicolumn{1}{l|}{94.72}     & 29\%            \\ \cline{3-11} 
\multicolumn{1}{|l|}{}                            & \multicolumn{1}{l|}{}                                    & \multicolumn{1}{l|}{32}                  & \multicolumn{1}{l|}{95.21}     & \multicolumn{1}{l|}{42914}   & \multicolumn{1}{l|}{90.09}      & \multicolumn{1}{l|}{85.31}     & \multicolumn{1}{l|}{22\%}             & \multicolumn{1}{l|}{96.61}     & \multicolumn{1}{l|}{90.26}     & 17\%            \\ \cline{3-11} 
\multicolumn{1}{|l|}{}                            & \multicolumn{1}{l|}{}                                    & \multicolumn{1}{l|}{37}                  & \multicolumn{1}{l|}{85.88}     & \multicolumn{1}{l|}{23198}   & \multicolumn{1}{l|}{89.19}      & \multicolumn{1}{l|}{80.68}     & \multicolumn{1}{l|}{14\%}             & \multicolumn{1}{l|}{96.61}     & \multicolumn{1}{l|}{81.51}     & 13\%            \\ \hline
\multicolumn{11}{|l|}{Low   delay configuration}                                                                                                                                                                                                                                                                                                                                                                       \\ \hline
\multicolumn{1}{|l|}{\multirow{4}{*}{PartyScene}} & \multicolumn{1}{l|}{\multirow{4}{*}{832 $\times$ 480}}   & \multicolumn{1}{l|}{22}                  & \multicolumn{1}{l|}{98.48}     & \multicolumn{1}{l|}{16641}   & \multicolumn{1}{l|}{92.03}      & \multicolumn{1}{l|}{81.52}     & \multicolumn{1}{l|}{35\%}             & \multicolumn{1}{l|}{92.77}     & \multicolumn{1}{l|}{90.87}     & 24\%            \\ \cline{3-11} 
\multicolumn{1}{|l|}{}                            & \multicolumn{1}{l|}{}                                    & \multicolumn{1}{l|}{27}                  & \multicolumn{1}{l|}{94.89}     & \multicolumn{1}{l|}{8306}    & \multicolumn{1}{l|}{86.65}      & \multicolumn{1}{l|}{79.89}     & \multicolumn{1}{l|}{34\%}             & \multicolumn{1}{l|}{92.77}     & \multicolumn{1}{l|}{87.91}     & 24\%            \\ \cline{3-11} 
\multicolumn{1}{|l|}{}                            & \multicolumn{1}{l|}{}                                    & \multicolumn{1}{l|}{32}                  & \multicolumn{1}{l|}{87.43}     & \multicolumn{1}{l|}{3811}    & \multicolumn{1}{l|}{84.13}      & \multicolumn{1}{l|}{75.92}     & \multicolumn{1}{l|}{27\%}             & \multicolumn{1}{l|}{92.77}     & \multicolumn{1}{l|}{81.14}     & 24\%            \\ \cline{3-11} 
\multicolumn{1}{|l|}{}                            & \multicolumn{1}{l|}{}                                    & \multicolumn{1}{l|}{37}                  & \multicolumn{1}{l|}{73.56}     & \multicolumn{1}{l|}{1609}    & \multicolumn{1}{l|}{83.17}      & \multicolumn{1}{l|}{67.37}     & \multicolumn{1}{l|}{10\%}             & \multicolumn{1}{l|}{92.77}     & \multicolumn{1}{l|}{67.99}     & 22\%            \\ \hline
\multicolumn{1}{|l|}{\multirow{4}{*}{BQTerrace}}  & \multicolumn{1}{l|}{\multirow{4}{*}{1920 $\times$ 1080}} & \multicolumn{1}{l|}{22}                  & \multicolumn{1}{l|}{99.08}     & \multicolumn{1}{l|}{113900}  & \multicolumn{1}{l|}{95.82}      & \multicolumn{1}{l|}{87.83}     & \multicolumn{1}{l|}{67\%}             & \multicolumn{1}{l|}{96.61}     & \multicolumn{1}{l|}{95.72}     & 53\%            \\ \cline{3-11} 
\multicolumn{1}{|l|}{}                            & \multicolumn{1}{l|}{}                                    & \multicolumn{1}{l|}{27}                  & \multicolumn{1}{l|}{97.76}     & \multicolumn{1}{l|}{29848}   & \multicolumn{1}{l|}{92.27}      & \multicolumn{1}{l|}{86.83}     & \multicolumn{1}{l|}{60\%}             & \multicolumn{1}{l|}{96.61}     & \multicolumn{1}{l|}{94.09}     & 49\%            \\ \cline{3-11} 
\multicolumn{1}{|l|}{}                            & \multicolumn{1}{l|}{}                                    & \multicolumn{1}{l|}{32}                  & \multicolumn{1}{l|}{94.25}     & \multicolumn{1}{l|}{7415}    & \multicolumn{1}{l|}{90.09}      & \multicolumn{1}{l|}{84.68}     & \multicolumn{1}{l|}{37\%}             & \multicolumn{1}{l|}{96.61}     & \multicolumn{1}{l|}{89.55}     & 28\%            \\ \cline{3-11} 
\multicolumn{1}{|l|}{}                            & \multicolumn{1}{l|}{}                                    & \multicolumn{1}{l|}{37}                  & \multicolumn{1}{l|}{85.19}     & \multicolumn{1}{l|}{2582}    & \multicolumn{1}{l|}{89.19}      & \multicolumn{1}{l|}{80.13}     & \multicolumn{1}{l|}{18\%}             & \multicolumn{1}{l|}{96.61}     & \multicolumn{1}{l|}{81.01}     & 18\%            \\ \hline
\end{tabular}

\end{table*}

\begin{table}[!ht]
\centering
\caption{Comparison of IQNet and PS~\cite{wang2023surprise}.}
\label{tab:comparison}
\resizebox{\columnwidth}{!}{%
\begin{tabular}{@{}llllllllll@{}}
\toprule
\multirow{3}{*}{Sequences} & \multirow{3}{*}{Qp} & \multicolumn{4}{c}{All intra} & \multicolumn{4}{c}{Low delay} \\ \cmidrule(lr){3-6} \cmidrule(lr){7-10}
                           &                     & \multicolumn{2}{c}{PSNR drop} & \multicolumn{2}{c}{Bitrate saving} & \multicolumn{2}{c}{PSNR drop} & \multicolumn{2}{c}{Bitrate saving} \\ \cmidrule(lr){3-4} \cmidrule(lr){5-6} \cmidrule(lr){7-8} \cmidrule(lr){9-10}
                           &                     & PS            & IQNET         & PS               & IQNET           & PS            & IQNET         & PS               & IQNET           \\  
\midrule
\multirow{4}{*}{BQTerrace} & 22                  & 6.08          & 5.14          & 50.42            & 31.52           & 2.43          & 3.59          & 61.76            & 52.59           \\
                           & 27                  & 2.18          & 2.17          & 25.99            & 29.29           & 0.38          & 1.50          & 21.93            & 49.00           \\
                           & 32                  & 1.12          & 1.15          & 15.55            & 17.49           & 0.19          & 0.97          & 10.04            & 27.98           \\
                           & 37                  & 1.05          & 0.82          & 15.29            & 13.22           & 0.1           & 0.72          & 5.94             & 18.36           \\
\midrule
\multirow{4}{*}{Cactus}    & 22                  & 3.57          & 4.30          & 49.23            & 32.97           & 1.35          & 3.37          & 49.05            & 51.68           \\
                           & 27                  & 0.27          & 2.49          & 14.44            & 21.11           & 0.34          & 2.16          & 13.06            & 26.11           \\
                           & 32                  & 0.86          & 1.69          & 12.48            & 14.61           & 0.31          & 1.51          & 9.63             & 14.10           \\
                           & 37                  & 0.87          & 1.12          & 12.54            & 11.14           & 0.29          & 1.02          & 8.86             & 9.86            \\
\midrule
\multirow{4}{*}{ParkScene} & 22                  & 4.22          & 2.69          & 47.13            & 19.88           & 1.86          & 2.17          & 33.52            & 25.68           \\
                           & 27                  & 1.77          & 1.68          & 24.43            & 17.24           & 0.46          & 1.42          & 12.66            & 19.74           \\
                           & 32                  & 0.94          & 1.01          & 15.93            & 14.86           & 0.35          & 0.88          & 10.93            & 15.47           \\
                           & 37                  & 0.86          & 0.57          & 17.05            & 11.57           & 0.28          & 0.51          & 10.07            & 11.28           \\
\midrule
\multirow{4}{*}{BQMall}    & 22                  & 4.7           & 2.97          & 39.58            & 17.58           & 2.16          & 2.42          & 33.32            & 24.28           \\
                           & 27                  & 2.12          & 1.98          & 20.81            & 14.11           & 0.55          & 1.67          & 11.44            & 17.46           \\
                           & 32                  & 1.2           & 1.31          & 13.32            & 12.23           & 0.42          & 1.12          & 9.3              & 13.87           \\
                           & 37                  & 1.07          & 0.84          & 13.31            & 10.05           & 0.31          & 0.75          & 7.08             & 10.60           \\
\midrule
\multirow{4}{*}{PartyScene}& 22                  & 5.95          & 4.21          & 39.64            & 17.18           & 3.2           & 4.52          & 37.61            & 24.47           \\
                           & 27                  & 1.62          & 2.83          & 14.84            & 16.73           & 0.77          & 2.79          & 14.12            & 24.35           \\
                           & 32                  & 1.43          & 2.02          & 14.97            & 17.68           & 0.5           & 1.76          & 11.21            & 24.16           \\
                           & 37                  & 1.21          & 1.43          & 15.34            & 17.51           & 0.29          & 1.06          & 8.93             & 21.53           \\

\midrule
\multirow{4}{*}{RaceHorses}& 22                  & 5.52          & 4.24          & 40.95            & 15.07           & 3.18          & 3.41          & 40.75            & 23.66           \\
                           & 27                  & 2.43          & 2.71          & 22.16            & 16.63           & 0.79          & 2.33          & 13.87            & 22.55           \\
                           & 32                  & 1.31          & 1.63          & 14.78            & 16.39           & 0.61          & 1.71          & 10.81            & 21.12           \\
                           & 37                  & 1.07          & 0.87          & 17.16            & 12.39           & 0.45          & 1.23          & 10.23            & 16.49           \\
\midrule                           
average                    &                     & 2.23          & 2.16          & 23.64            & 17.44           & 0.898         & 1.86          & 19.05            & 23.6            \\ 
\bottomrule
\end{tabular}%
}

\end{table}

\subsection{The visual results of the IQNet}

Fig.~\ref{comparision_face} shows a detailed visual comparison between the reconstructed result of the first frame of $BQterrace$ with and without IQNet prefiltering for the all-intra encoding at QP27. As depicted in Fig.~\ref{comparision_face} (b) and (c), the proposed one on the girl's face exhibits higher perceptual quality due to our filtering, which avoids the artifacts in the reconstructed anchor image such as the noise on the face. IQNet could filter this unnecessary information to achieve better quality. Similarly, for other areas such as the background (chair and man's back) in Fig.~\ref{comparision_face} (d) and (e), IQNet could reduce details with a small drop in perceptual quality for bitrate reduction. The reduction in bitrate of the $BQterrace$ sequence in QP27 is about 30\% with IQNet for the all-intra encoding.

Fig.~\ref{horse_result32} presents another comparison at QP32. We find that the perceptual quality of these two is similar, even though IQNet is trained at QP27.
From this figure, we can see that the perceptual quality of the prefiltered reconstructed image in saliency regions such as humans or animals is similar to the anchor. This is because the prediction of the JND scale is similar regardless of the QP used for reconstructed image production in training data generation. However, the JND scale selected at QP27 is more distinguishable than the ones under other QPs in our experience, which is also the reason why we chose it as the base QP. As a result, although IQNet is trained at a specific QP, it can be applied directly to prefiltering at other QPs.

To demonstrate how IQNet learns from the training data, we randomly select two images and apply IQNet along with the proposed training data generation flow. Table~\ref{table_gt} displays the bitrate and the VMAF~\cite{vmaf} rating. VMAF predicts subjective quality by combining multiple elementary quality metrics through machine learning. As depicted in the table, both exhibit similar VMAF ratings. With appropriate learning, our IQNet is well-trained to offer a performance akin to that of the intricate training data generation.

\subsection{The results of the bitrate, PSNR, and VMAF}

Fig.~\ref{vmaf_result} illustrates the video quality comparison between anchors and our IQNet prefiltering in terms of VMAF for all-intra and low-delay P configurations at varying QPs. The Y-axis in Fig.~\ref{vmaf_result} to \ref{bitrate_result} is arranged in the order from top to bottom as shown in Table~\ref{HEVC seq}. As depicted in these figures, the VMAF ratings remain consistent for both scenarios irrespective of the QPs and encoding profiles, indicating comparable quality. This also confirms the subjective quality comparison discussed in the preceding subsection.

Fig.~\ref{psnr_result} and Fig.~\ref{bitrate_result} show the PSNR drop and the bitrate reduction with IQNet prefiltering. Regarding the contribution to bitrate saving, since our prefiltering network could effectively find perceptual redundancy in the input sequence and remove them before encoding, it can achieve maximum 41\%, minimum 5\% and average 15\% bitrate reductions using all-intra configuration in VVC, and maximum 53\%, minimum 4\%, and average 19\% bitrate reductions using low-delay configuration P in VVC for all 18 HEVC test sequences. In addition, we observe that the bitrate reduction tends to decrease with increasing QP. The reason might be that the quantization effect would be larger at higher QPs and the perceptual redundancy in images would eventually disappear whether we remove them before encoding or not. As a result, the effect of prefiltering would be lower at high QP. Regarding the PSNR drop, IQNet has higher PSNR drops at lower QP. For VMAF drop, IQNet has two kinds of results: higher drop at lower QPs as PSNR or lower drop at lower QPs. One possible reason for the lower VMAF drop at lower QPs could be due to the well preserved edges, which leads to higher quality but also lower bitrate saving.

\begin{figure}[tb]
    \centering
     \begin{subfigure}[b]{0.8\linewidth}
     \centering
     \includegraphics[width=\linewidth]{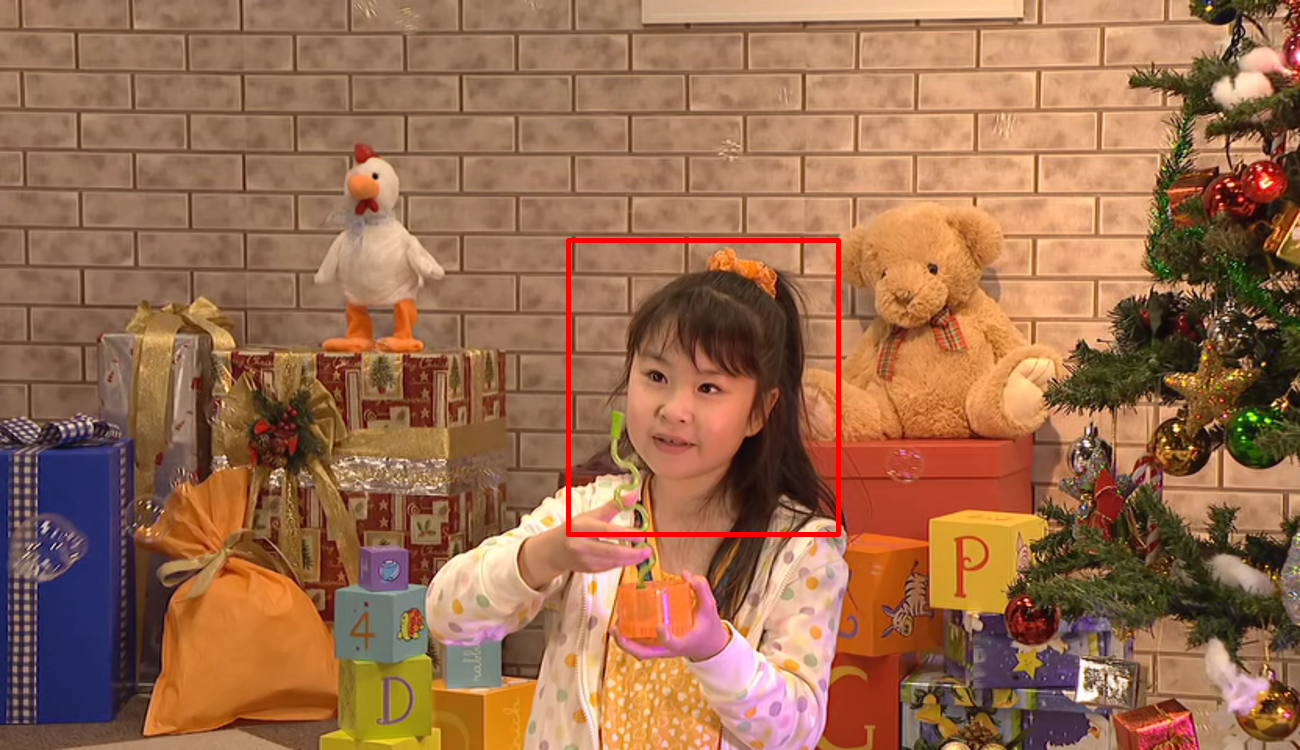}
     \caption{anchor}
     \label{horsespot444}
     \end{subfigure}    
     \begin{subfigure}[b]{0.32\linewidth}
         \centering
         \includegraphics[width=\linewidth]{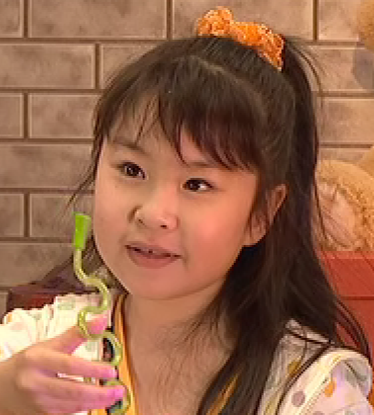}
         \caption{anchor}
         \label{woman_ori_rec32}
     \end{subfigure}
     \hfill
     \begin{subfigure}[b]{0.32\linewidth}
         \centering
         \includegraphics[width=\linewidth]{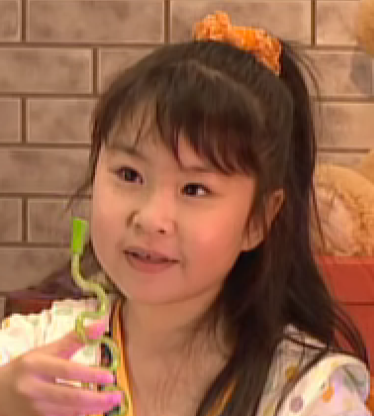}
         \caption{CNN-JNQD~\cite{ki2018erjnd}}
         \label{woman_iq_rec32}
     \end{subfigure}     
     \hfill
     \begin{subfigure}[b]{0.32\linewidth}
         \centering
         \includegraphics[width=\linewidth]{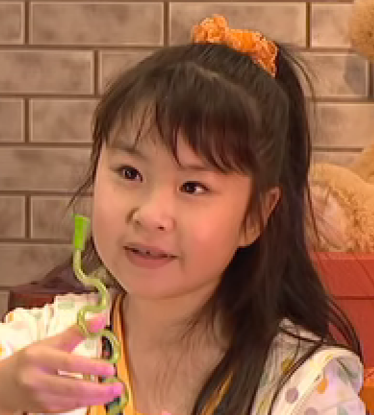}
         \caption{IQNet}
         \label{woman_cnn_rec32}
     \end{subfigure}
     \caption{The reconstructed frame of $PartyScene$ encoded @QP22. (PSNR, bitrate) for the anchor, CNN-JNQD and IQNet are (40.97, 42643), (30.65, 28378), and (35.48, 34423), respectively. }
        \label{party_triple}
\end{figure}

\subsection{The comparison with other works}

To compare our IQNet with CNN-JNQD~\cite{ki2018erjnd}, we arbitrarily select two test sequences, $PartyScene$ and $BQTerrace$, and assess their VMAF rating and bitrate as shown in Table~\ref{table_iqcnn3}. From Table~\ref{table_iqcnn3}, it is discernible that our IQNet achieves a higher VMAF rating compared to CNN-JNQD for all prefiltered and reconstructed sequences in varying QPs. Nevertheless, in the aspect of bitrate reduction comparison, it emerges that our IQNet secures slightly lesser savings than CNN-JNQD, barring at high QP. The primary rationale is that our IQA criteria during training data generation incline towards retaining more details to better align with JND, albeit at the expense of lesser bitrate reduction. In contrast to CNN-JNQD, the savings in the BD-rate under VMAF for PartyScene and BQTerrace are (all intra: 33. 3\%/35. 2\%) and (low delay P: 43. 7\%/ 48\%), respectively. As Fig.~\ref{party_triple} illustrates, the girl in the CNN-JNQD image appears more blurry (specifically the hair and eyes) compared to the anchor and our result. CNN-JNQD exhibits markedly lower quality and does not satisfy the JND requirement.
Our IQNet endeavors to preserve more details to attain superior quality, particularly in the human face, although this approach results in lesser bitrate reduction compared to CNN-JNQD.
However, for a comparable VMAF rating like $PartyScene$ at QP37, our methodology could economize on more bitrates than CNN-JNQD.

Besides the comparison of bitrate and VMAF, it is noteworthy that CNN-JNQD necessitates a unique model for each QP, while our IQNet merely requires a singular model for all QPs, rendering it more user-friendly. Additionally, our model size is significantly smaller than that of CNN-JNQD, being an order of magnitude smaller.

Table~\ref{tab:comparison} shows the comparison with PS~\cite{wang2023surprise}. The average bitrate reduction of the IQNet is larger than that of the PS with slightly higher PSNR drop for the low delay case due to the our leanring based approach. In contrast, the average bitrate reduction of IQNet is smaller than that of the PS with similar average PSNR drop for the all intra case due to our details preservation approach. 

\section{Conclusion}

Addressing the problems of the time-consuming and complex JND modeling, this paper proposes a no-reference IQA-guided JND prefiltering network to determine JND through a scalable and systematic approach. The network is trained on our proposed fine-grained JND dataset, constructed from decoded images to include coding effects, and perceptually enhanced with block overlap and edge preservation. Furthermore, the lightweight JND prefiltering network, IQNet, removes image redundancy before encoding. Developed on a base QP, it is applicable to different QPs and requires only 3K parameters. The experimental results show that our network could achieve a maximum 41\%, minimum 5\% and average 15\% bitrate reduction of all test sequences using the all-intra configuration in VVC. Meanwhile, a maximum 53\%, minimum 4\% and average 19\% bitrate reduction was observed using the low-delay P configuration in VVC, with negligible subjective quality loss. Compared to the previous CNN-JNQD, which required expensive subjective tests, our method achieves much higher subjective quality without blurry images, while maintaining similar bitrate savings. For future work, employing an IQA other than NIMA and a JND model other than ERJND could better model real JND with a larger dataset. In addition, JND on chrominance can be included as well. The proposed approach can also be adapted to JND in quantization for higher coding efficiency.

\bibliographystyle{IEEEtran}

\bibliography{bib/ieeeBSTcontrol,bib/thesis}

\begin{IEEEbiography}[{\includegraphics[width=1in,height=1.25in,clip,keepaspectratio]{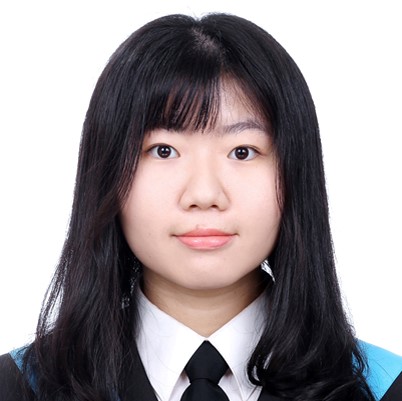}}]{Yu-Han Sun}
received the M.S. degree in electronics engineering from the National Yang Ming Chiao Tung University, Hsinchu, Taiwan, in 2022. She is currently working in the Mediatek, Hsinchu, Taiwan. Her research interest includes perceptual image processing and VLSI design.

\end{IEEEbiography}

\begin{IEEEbiography}[{\includegraphics[width=1in,height=1.25in,clip,keepaspectratio]{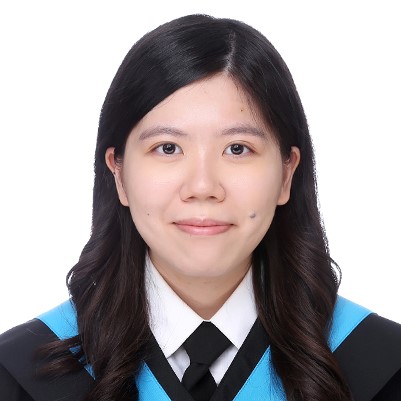}}]{Chiang Lo-Hsuan Lee}
is currently pursuing the M.S. degree in electronics engineering from the National Yang Ming Chiao Tung University, Hsinchu, Taiwan, in 2022. Her research interest includes perceptual image processiong and VLSI design.

\end{IEEEbiography}

\begin{IEEEbiography}[{\includegraphics[width=1in,height=1.25in,clip,keepaspectratio]{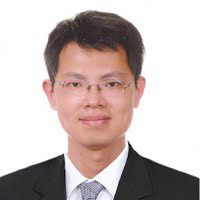}}]{Tian-Sheuan Chang}
	(S’93–M’06–SM’07)
	received the B.S., M.S., and Ph.D. degrees in electronic engineering from National Chiao-Tung University (NCTU), Hsinchu, Taiwan, in 1993, 1995, and 1999, respectively. 
	
	From 2000 to 2004, he was a Deputy Manager with Global Unichip Corporation, Hsinchu, Taiwan. In 2004, he joined the Department of Electronics Engineering, NCTU (as National Yang Ming Chiao Tung University (NYCU) in 2021), where he is currently a Professor. In 2009, he was a visiting scholar in IMEC, Belgium. His current research interests include system-on-a-chip design, VLSI signal processing, and computer architecture.
	
	Dr. Chang has received the Excellent Young Electrical Engineer from Chinese Institute of Electrical Engineering in 2007, and the Outstanding Young Scholar from Taiwan IC Design Society in 2010. He has been actively involved in many international conferences as an organizing committee or technical program committee member.
\end{IEEEbiography}
\end{document}